\documentstyle[preprint,aps,prb]{revtex}

\begin{document}
\draft
\preprint{}
\title{Magic Numbers for Classical Lennard-Jones Cluster \\
    Heat Capacities}
\author{D. D. Frantz}
\address{Department of Chemistry, University of Lethbridge, \\
    Lethbridge, T1K 3M4, Canada}
\date{\today}
\maketitle
\begin{abstract}
Heat capacity curves as functions of temperature for classical atomic
clusters bound by pairwise Lennard-Jones potentials were calculated for
aggregate sizes $4 \leq N \leq 24$ using Monte Carlo methods.
J-walking (or jump-walking) was used to overcome convergence
difficulties due to quasi-ergodicity in the solid-liquid transition
region. The heat capacity curves were found to differ markedly and
nonmonotonically as functions of cluster size. Curves for $N = 4$, 5
and 8 consisted of a smooth, featureless, monotonic increase throughout
the transition region, while curves for $N = 7$ and 15--17 showed a
distinct shoulder in this region; the remaining clusters had
distinguishable transition heat capacity peaks.  The size and location
of these peaks exhibited ``magic number'' behavior, with the most
pronounced peaks occurring for magic number sizes of $N = 13$, 19 and
23. This is consistent with the magic numbers found for many other
cluster properties, but there are interesting differences for some of
the other cluster sizes.  Further insight into the transition region
was obtained by comparing rms bond length fluctuation behavior with the
heat capacity trends.  A comparison of the heat capacities with other
cluster properties in the solid-liquid transition region that have been
reported in the literature indicates partial support for the view that,
for some clusters, the solid-liquid transition region is a coexistence
region demarcated by relatively sharp, but separate, melting and
freezing temperatures; some discrepancies, however, remain unresolved.
\end{abstract}

\narrowtext

\section{Introduction}
The investigation of small atomic and molecular clusters is an active
area of current research, both
%% FOLLOWING LINE CANNOT BE BROKEN BEFORE 80 CHAR
theoretically\cite{coexist,WB,BJB,BBDJ,HMD,BM,AS,AS2,BJ,GBB,HA,QS,EK,EK1,BB,LBA}
and experimentally.\cite{ELR,FFRT,ESR} Their small size (typically 3 to
150 atoms) makes clusters ideal candidates for computer simulation
studies, and most theoretical studies to date have used Monte
Carlo\cite{MRT2,Val-Whit,Kalos-Whit,Allen-Tild} and molecular dynamics
methods.\cite{Allen-Tild,Haile,Hockney-Eastwood} Because of the large
computational demands inherent in these simulation methods, most early
theoretical investigations were limited to classical rare gas clusters
bound by simple pairwise additive Lennard-Jones potentials, but
continuing improvements in computer technology have led to an
increasing number of more interesting metal cluster simulations
requiring more sophisticated intermolecular potentials for physically
realistic representation,\cite{VSA} and have made quantum simulations
based on Fourier path integral methods practical.\cite{FPI_rev}

One major motivation for the study of clusters is the insight that can
be provided for understanding the transition from finite to bulk
behavior. Many cluster properties show a very irregular dependence on
the aggregate cluster size $N$. For example, the difference in the
binding energy between successive pairs of rare gas clusters is
especially marked for clusters of size $N = 13$, 19, 23, 26, 29, 39,
46, 49, and 55, and dramatically smaller for their immediate
neighbors.\cite{Northby} The mass spectrum intensity of Xe clusters
formed by condensation in supersonic jets was found to increase
slightly for sizes $N = 13$, 19, 25, 55, 71, 87, and 147 and then
decrease sharply for the clusters immediately following in
size.\cite{ESR} Calculations of the Gibbs free energy of formation for
argon clusters with sizes ranging from 2 to 20 indicated local minima
for $N = 7$, 13 and 19.\cite{FD_magic} This nonuniform dependence of
certain cluster properties on size, often termed ``magic number''
effects, has been of great theoretical interest. Much work has been
done in relating magic number sequences to cluster structure in terms
of ``soft'' sphere packings, since many magic number sizes such as $N =
13$, 19, 55, and 147 correspond to compact icosahedral based
structures.\cite{Northby,Hoare-Pal,MF}

An especially important phenomenon exhibiting magic number effects is
cluster ``melting.'' Because of their small size, clusters do not have
the sharp first order transition characteristic of bulk melting, but
instead show transitions occurring over a range of
temperatures.\cite{BBDJ} The transition range generally decreases as
the cluster size increases, coalescing to the bulk transition
temperature as $N$ approaches infinity, but it shows a wide variety of
behavior for smaller sizes. Berry and coworkers\cite{coexist} have
postulated that for certain sized clusters, the transition region is in
fact a coexistence region where both ``solid-like'' and ``liquid-like''
isomers dynamically coexist, fluctuating back and forth between
relatively long lived ``solid'' and ``liquid'' states. This implies
that these clusters have sharp, but different, ``freezing'' and
``melting'' temperatures. Some of their molecular dynamics studies
revealed that clusters with sizes $N = 7$, 9, 11, 13, 15, and 19 have
especially high melting temperatures and pronounced coexistence ranges,
while clusters with sizes $N = 8$, 14 and 20 melt at much lower
temperatures and exhibit no coexistence.\cite{BBDJ}

The problem of cluster phase transitions, has been one of great
interest, and one not lacking in controversy (Ref.~\onlinecite{BBDJ}
provides a good review). While many studies are consistent with the
view of a dynamic coexistence between solid-like and liquid-like
isomers in the transition region, other studies emphasize, instead, a
relatively smooth progression of isomerizations between locally similar
but globally distinct configurations occurring throughout the
transition region.\cite{AS} In addition, it is becoming increasingly
clear that the problem of incomplete sampling of configuration space
for simulations run in the transition region is much more serious than
had been appreciated earlier, and that extremely long simulations (much
longer than those routinely used even a few years ago) are necessary to
prevent large systematic errors in certain properties that can result
in misleading interpretations. Given the large amount of contradictory
evidence that has been accumulated over the years, as well as the
problem of not knowing how reliable some of it is, it is apparent that
the nature of cluster melting remains still very much unresolved.

There is no problem concerning cluster behavior at extreme
temperatures. At very low temperatures, clusters are locked into solid
structures having near rigid rotations and small amplitude,
harmonic-like vibrations; there is no conversion between different
isomers. At sufficiently high temperatures, the clusters dissociate,
implying that for simulations run under free volume conditions, the
average energy vanishes in the limit of an infinite number of
configurations.\cite{QS,EK,EK1} Consequently, it is usual in Monte
Carlo simulations to confine the cluster within a perfectly reflecting
constraining potential having a typical radius of a few atomic radii
that is centered on the cluster's center of mass.\cite{LBA} Under such
conditions, the high temperature limit corresponds then to a highly
compressed vapor confined to a small spherical cavity. This limiting
behavior is evident in Fig.~\ref{Fig:Ar23}, which shows the average
internal energy and heat capacity as functions of temperature for a
typical small classical rare gas cluster. For $T \rightarrow 0$, the
energy approaches the global minimum potential energy, and the heat
capacity approaches the classical equipartition of energy result for a
harmonically bound nonlinear system, $3N - 3$.  For $T \rightarrow
\infty$, the internal energy and heat capacity both approach their
ideal gas limits.

What happens in between is less clear. In this case, the heat capacity
curve has two distinct peaks, which would indicate two ``phase''
changes. Temperature driven first-order transitions in small systems
are typically spread out over a finite range of temperatures, with the
size of the transition region corresponding to the width of the heat
capacity peak, and the transition temperature corresponding to the heat
capacity maximum.\cite{BBDJ} The larger, higher temperature heat
capacity peak can be therefore associated with the transition from a
condensed phase to the vapor phase (that is, cluster dissociation),
while the smaller, lower temperature peak represents a solid-liquid
transition. Not surprisingly, the size and location of the higher
temperature peak have a sensitive dependence on the cluster simulation
boundary conditions, and thus vary considerably as functions of the
constraining radius. Their dependence on the cluster aggregate size is
less pronounced, though,  and quite regular.  Hence the information
conveyed in this region is somewhat artificial and lacking in physical
importance. This is not the case for the solid-liquid transition,
however. This peak is largely independent of the constraining radius,
provided the radius is not so small that the free rearrangement of the
cluster is limited. In addition, the solid-liquid transition region has
a very marked dependence on cluster size; while all clusters confined
in sufficiently large constraining spheres show a large dissociation
peak in their heat capacity-temperature curve, not all clusters show
peaks in the solid-liquid transition region, and for those that do, the
size and shape of the peak have a very nonuniform dependence on cluster
size.  Does the presence of a heat capacity peak in the solid-liquid
transition region then necessarily imply a solid-liquid transition
temperature corresponding to the peak temperature, and concomitantly,
does the absence of a heat capacity peak there rule out a definite
solid-liquid transition temperature? What is the connection between the
transition region as defined by the heat capacity peak width and the
coexistence region defined by the sharp, yet distinct, freezing and
melting temperatures obtained from some molecular dynamics
simulations?  One would expect the two ranges to mostly coincide, but
as will be seen, this is not always the case.

Despite the importance of heat capacities in characterizing cluster
phase transitions, few cluster heat capacity studies have been reported
in the literature, and most of these have been limited to special
cases, particularly $N = 13$\cite{QS,Tsai-Jordan2,DJB,LW} and $N =
55$.\cite{LW,NP} This is due to the notorious difficulty in extracting
heat capacities with sufficient accuracy from simulations.  A typical
example of the poor convergence of heat capacities is given in
Fig.~\ref{Fig:Ar23}. The heat capacity curve obtained from averages
taken over $10^5$ Monte Carlo passes (or $23 \times 10^5$ moves) for
each temperature point (a simulation length that was typical ten to
fifteen years ago) is not even qualitatively correct, with the
solid-liquid transition peak completely absent. The curve obtained from
walk lengths of $10^6$ passes (a length routinely achieved five years
ago) does show a small solid-liquid transition peak, but one that is
very much diminished. Even the curve based on $10^7$ passes is
quantitatively poor in this region.  This poor convergence behavior is
a result of the incomplete sampling of configuration space by the
random walker, a phenomenon denoted as ``quasi-ergodicity'' by Valleau
and Whittington.\cite{Val-Whit} Quasi-ergodicity arises in systems
where the sample space contains two or more regions that have a very
low transition probability, resulting in bottlenecks that effectively
confine the sampling to only one of the regions.\cite{Allen-Tild} In
clusters such as Ar$_{23}$, the potential hypersurface consists of very
deep wells, corresponding to stable solid isomers, separated by very
large barriers. The resultant dichotomy of time scales characterizing
the Monte Carlo random walks (or likewise, the molecular dynamics time
evolution) produces rapid small-scale motion within the wells, but very
slow large-scale movement between them.\cite{Cao-Berne} Because of the
finite length of the walks (or trajectories), systematic errors result
that diminish only with increasing length, disappearing in the limit of
infinitely long walks (or trajectories). Interestingly, it is precisely
these conditions that Berry and coworkers claim are necessary for the
existence of a solid-liquid coexistence region in clusters;\cite{BBDJ}
it appears that the very nature of this region that makes it so
physically intriguing also conspires to make its elucidation
exceedingly difficult.

Recently, new methods have been developed that substantially reduce
quasi-ergodicity and dramatically increase convergence in cluster Monte
Carlo simulations in the transition regions, thus allowing even
computationally difficult properties such as cluster heat capacities to
be obtained to high accuracy. These methods are characterized by their
emphasis on large scale motions, or jumps, of the random walker to
other regions of configuration space and thus have been collectively
termed ``J-walking'' (for jump-walking).\cite{J-walker} J-walking has
been successfully applied to several systems, including binary metal
clusters,\cite{Lopez-Freeman} water clusters,\cite{Tsai-Jordan} and
clusters absorbed on surfaces.\cite{SLF} Tsai and
Jordan\cite{Tsai-Jordan2,Tsai-Jordan} have combined J-walking with
histogram methods,\cite{Ferr-Swend} and recently, J-walking was
extended to quantum systems.\cite{JWFPI} With J-walking, it is now
feasible to systematically study the dependence of the heat capacity on
cluster size for a fairly large range of cluster sizes; the solid
curves in Fig.~\ref{Fig:Ar23} were obtained using J-walking.

I begin in Section~\ref{Sec:theory} with a brief review of the
J-walking method and a summary of the calculations undertaken.
Section~\ref{Sec:results} describes the results obtained using
J-walking for the heat capacities and potential energies as functions
of temperature and cluster size for Lennard-Jones clusters ranging in
size from 4 to 24 atoms. Also included are results obtained from
similar standard Monte Carlo simulations, including rms bond length
flucuations, as well as comparisons with molecular dynamics simulations
and other calculations reported in the literature for rare gas clusters
in this range. Finally, in Section~\ref{Sec:discuss}, I discuss my
findings in the context of some of the controversies and ambiguities
concerning cluster phase transitions that remain unresolved.

\section{Method}                        \label{Sec:theory}
J-walking addresses the problem of quasi-ergodicity in standard Monte
Carlo simulations based on the sampling algorithm proposed by
Metropolis {\em et al.},\cite{MRT2} hereinafter called Metropolis Monte
Carlo, by coupling the usual small-scale Metropolis moves with
occasional large-scale jumps that move the random walker to different
regions of configuration space. For classical systems governed by a
potential $V({\bf r})$, the Metropolis method uses a random walk to
sample configuration space, with each attempted move from the current
location ${\bf r}_i$ to a trial location ${\bf r}_f$ determined by the
acceptance probability,
\begin{equation}
    P = \min\{1,q({\bf r}_f|{\bf r}_i)\},
\end{equation}
where
\begin{equation}
    q({\bf r}_f|{\bf r}_i) = \frac{T({\bf r}_i|{\bf r}_f)\rho({\bf r}_f)}
        {T({\bf r}_f|{\bf r}_i)\rho({\bf r}_i)},
    \label{qdef}
\end{equation}
is the acceptance ratio, $\rho({\bf r})=Z^{-1}\exp\{-\beta V({\bf
r})\}$ is the Boltzmann distribution with $Z$ the standard
configuration integral and $\beta = 1/k_BT$ the temperature parameter,
and $T({\bf r}'|{\bf r})$ is the transition matrix or sampling
distribution. The sampling distribution is usually generated from
uniform deviates $\xi$ over a finite stepsize range $\Delta$ to
give\cite{Kalos-Whit}
\begin{eqnarray}
    T({\bf r}_f|{\bf r}_i)  & = & \left\{
        \begin{array}{cl}
            1/\Delta  & \mbox{for\ \ } |{\bf r}_f - {\bf r}_i|
            < \Delta/2, \\
             0 & \mbox{otherwise}, \\
        \end{array} \right.                     \label{TUD}
\end{eqnarray}
and thus
\begin{equation}
    q({\bf r}_f|{\bf r}_i) = \exp\{-\beta[V({\bf r}_f)-V({\bf r}_i)]\}.
    \label{qMRT2}
\end{equation}
Attempted moves are generated according to ${\bf r}_f = {\bf r}_i +
(\xi - \frac{1}{2})\Delta$, with the maximum stepsize $\Delta/2$ usually
adjusted to give acceptance probabilities of approximately 50\%. The
required size depends on the width of the Boltzmann distribution, and
thus decreases with increasing $\beta$. This temperature dependence can
lead to quasi-ergodic behavior whenever the stepsize becomes too small
relative to the potential barrier heights and widths; the walker
becomes effectively trapped within a region of configuration space if
too many small steps are required to surmount large, steep barriers
within the duration of the walk. Thus, for a given walk length, there
is a threshold temperature such that walks undertaken below that
temperature are quasi-ergodic.\cite{J-walker}

The Boltzmann distribution's dependence on temperature that can lead to
quasi-ergodicity can also be exploited to overcome it.  Although the
form of the Boltzmann distribution is largely dependent on the form of
the underlying potential, with distribution maxima corresponding to
potential minima, the widths of the distribution peaks have an inverse
dependence on $\beta$, with higher temperatures resulting in wider
distributions. In essence, higher temperature walkers are less
constrained because their larger stepsizes allow them to overcome the
potential barriers more effectively. Hence, a Boltzmann distribution
generated at some higher temperature where the sampling is ergodic can
be used by a lower temperature walker to make large-scale moves in
configuration space.  J-walking then occasionally replaces the usual
attempted Metropolis moves with attempted jumps to positions occupied
by such a higher temperature walker (J-walker). Because the peaks in
the Boltzmann distribution correspond to the potential minima, the
J-walker's motion remains biased about the minima, greatly increasing
the likelihood an attempted jump would be accepted.

This scheme is equivalent to replacing the usual Metropolis transition
matrix given by Eq.~(\ref{TUD}) with the Boltzmann distribution at the
higher temperature whenever a jump is attempted,
\begin{equation}
    T_J({\bf r}_J|{\bf r}_i) = Z^{-1}\exp\{-\beta_J V({\bf r}_J)\}
        \mbox{\ \ for\ \ } 0 \leq \xi_J < P_J,
\end{equation}
where $P_J$ is the jump attempt probability, $\beta_J$ is the J-walker
temperature parameter, and ${\bf r}_J$ is the J-walker's location. This
gives an acceptance ratio of
\begin{equation}
    q_J({\bf r}_J|{\bf r}_i) = \exp\{(\beta_J - \beta)
        [V({\bf r}_J) - V({\bf r}_i)]\}.        \label{qJwalker}
\end{equation}
The likelihood of an attempted jump being accepted depends on the two
Boltzmann distributions' overlap, which is reflected by the value of
$q_J$. In the high temperature limit $\beta_J \rightarrow 0$, the
acceptance ratio reduces to the standard Metropolis expression given in
Eq.~(\ref{qMRT2}).  Because the Boltzmann distribution broadens as the
temperature increases, J-walking in this limit reduces to
indiscriminate jumping with a very large stepsize, and so the
probability of a jump being accepted becomes very small. In the limit
$\beta_J \rightarrow \beta$, $q_J({\bf r}_J|{\bf r}_i) \rightarrow 1$
since the low temperature walker is now effectively sampling from its
own distribution.

Two complementary implementations for generating the classical J-walker
Boltzmann distributions were originally presented.\cite{J-walker} The
two are another example of the commonly encountered trade-off between
computer memory and speed. The first ran the J-walker in tandem with
the low temperature walker, with the low temperature walker
occasionally attempting jumps to the current J-walker position simply
by using the current J-walker coordinates as its trial position. In the
second implementation, the J-walker was run beforehand and the
configurations generated during the walk stored periodically in an
external array. Subsequent jump attempts were made by accessing the
stored configurations via randomly generated indices. The tandem walker
scheme has very little memory overhead, but has the disadvantage of
requiring that the J-walker be moved an extra number of steps whenever
a jump is attempted in order to reduce correlations between the two
walkers that otherwise result in systematic errors. The number of extra
steps needed depends on the temperature difference between the two
walkers, increasing the computation time greatly as the difference
becomes larger. This implementation then is more suited for parallel
computers. On the other hand, the use of external distributions has
only a modest computational overhead (mostly the time required to
generate the distributions), but requires very large storage facilities
for handling the distribution arrays.  Because fast workstations having
tens of Mb of RAM and Gb of disk storage are now affordable and quite
common, while access to parallel computers is still much more limited,
the implementation using externally stored distributions remains the
method of choice and was the one used for all the J-walking
calculations reported in this study.

J-walking simulations were run for all clusters in the range $4 \leq N
\leq 24$. The clusters studied were modeled by the usual pairwise
additive Lennard-Jones potential,
\begin{eqnarray}
    V & = & \sum_{i < j} V_{LJ}(r_{ij}), \nonumber \\
    V_{LJ}(r_{ij}) & = & 4\epsilon\left[\left(
        \frac{\sigma}{r_{ij}}\right)^{12}
        - \left(\frac{\sigma}{r_{ij}}\right)^6 \right].  \end{eqnarray}
Small clusters are known to become unstable beyond a threshold
temperature that varies with the cluster size.\cite{QS,EK1} For the
Lennard-Jones potential under free volume conditions, the average
energy vanishes in the limit of infinitely long walks.  Consequently,
the choice of boundary conditions can greatly affect some cluster
properties at higher temperatures.\cite{BBDJ} I have followed Lee,
Barker and Abraham\cite{LBA} and have confined the clusters by a
perfectly reflecting constraining potential of radius $R_c$ centered on
the cluster's center of mass. To maintain a common set of boundary
conditions throughout the survey, the constraining radius was identical
for all the clusters studied, with $R_c = 4\sigma$.

For physically realistic systems such as clusters, the Boltzmann
distributions are too narrow for a single distribution to be used for
J-walking over the entire temperature domain from the solid region to
the dissociation region, and so the distributions had to be generated
in stages. For each cluster, an initial J-walker distribution was
generated from a long Metropolis walk at a temperature high enough for
the sampling to be ergodic. This distribution was then used for
J-walking runs to obtain averages of the potential energy and the heat
capacity for a series of lower temperatures. The width of the J-walker
distribution limited the effective temperature range for subsequent
J-walking simulations. When the temperature difference between the
distribution and the low temperature walker became too large, very few
attempted jumps were accepted since the low J-walker configurations
likely to be accepted were in the low energy tail of the distribution.
At the temperature where the jump acceptance started becoming too
small,\cite{acceptance} a new distribution was generated from the
previous one, using J-walking to ensure it was ergodic as well. This
distribution was then used to obtain data for the next lower
temperature range, and then to generate the next lower temperature
J-walker distribution, and so on, until the entire temperature domain
was spanned. To ensure that the higher temperature J-walker
distributions were free of any systematic errors large enough to
adversely affect the sampling, the temperature ranges were occasionally
overlapped so that J-walking data collected near the end of one
distribution's useful temperature range (as indicated by the low jump
acceptance) could be compared with the data obtained from the beginning
of the next distribution's useful temperature range (where the jump
acceptance rate was high). For example, a J-walker distribution
generated at $T = 15$ K could be used to generate data over the
temperature range $11 \leq T \leq 15$ K. By generating the next
distribution at $T = 12$ K for use over the range $8 \leq T \leq 12$ K,
data for the overlapping range $11 \leq T \leq 12$ K obtained from both
distributions could be compared and checked for consistency.  All the
distributions checked in this manner were found to be consistent.

The classical internal energy and heat capacity were calculated by the
usual expressions for an $N$-atom cluster (in reduced units),
\begin{eqnarray}
    \left\langle U^* \right\rangle & = & \frac{3NT^*}{2}
        + \left\langle V^* \right\rangle, \\
    \left\langle C_V^* \right\rangle & = & \frac{3N}{2}
    + \frac{\left\langle{(V^*)^2}\right\rangle
    - \left\langle{V^*}\right\rangle^2}{(T^*)^2},
\end{eqnarray}
where $U^* = U/\epsilon$, $V^* = V/\epsilon$, $C_V^* = C_V/k_B$, and
$T^* = k_BT/\epsilon$. In order to more easily compare the classical
results with quantum simulations to be undertaken later, curves for
$\left\langle{U^*}\right\rangle$ and $\left\langle{C_V^*}\right\rangle$
as functions of $T^*$ were actually generated using Ne parameters, with
$\sigma = 2.749$~\AA\ and $\epsilon = 35.6$~K. The temperature mesh size
was typically $\Delta T = 0.02$~K, and was reduced to 0.01~K for the
solid-liquid transition regions. Similarly, J-walker distributions were
generated at Ne temperatures. All the J-walking simulations run for
each temperature consisted of $10^5$ warmup passes followed by $10^7$
passes with data accumulation (100 walks of $10^5$ passes each); the
jump attempt frequency was $P_J = 0.1$ throughout.

Table~\ref{Tbl:J-walker1} lists the J-walker distribution particulars
for clusters in the range $4 \leq N \leq 12$, while
Table~\ref{Tbl:J-walker2} lists them for the range $13 \leq N \leq
24$.  For each cluster, the initial J-walker distributions were
obtained at temperatures corresponding to the high temperature side of
the cluster dissociation peak, where all the cluster configurations
were completely fluid. In the original J-walking study for $N =
13$,\cite{J-walker} the initial J-walker distribution was obtained at a
temperature corresponding to the liquid region (just past the
solid-liquid transition region) where the sampling was thought to be
ergodic, but in fact turned out not to be fully ergodic. When J-walking
was extended to quantum systems and tested on $N = 7$
clusters,\cite{JWFPI} it was discovered that there was also much
quasi-ergodicity in the region corresponding to the low temperature
side of dissociation peak, and so it was necessary to begin the
temperature scans at an even higher temperature. This region
corresponds to a coexistence region where both liquid-like and
partially dissociated configurations are in dynamic coexistence, and
sampling difficulties can arise when single atoms leave the cluster and
wander about nearby for several passes before recombining with the
cluster.  Effects of quasi-ergodicity in this region can be seen in the
heat capacity curves for Ar$_{23}$ in Fig.~\ref{Fig:Ar23}.  On the high
temperature side of the dissociation peak ($T \gtrsim 80$ K) the
J-walker curve coincides with the Metropolis curves for both $10^6$ and
$10^7$ passes, while on the low temperature side of the peak ($50$ K
$\lesssim T \lesssim 75$ K ) there are substantial discrepancies
between the J-walking curve and the Metropolis curve obtained from
$10^6$ passes, and smaller discrepancies between the J-walking curve
and the Metropolis curve obtained from $10^7$ passes.  Although these
discrepancies are much smaller than those appearing in the solid-liquid
transition region, it is important to minimize systematic errors in the
J-walker distribution since the sensitivity of these errors on
subsequent J-walking runs is not known.

All J-walker distributions consisted of $10^6$ configurations, except
for $N = 4$ and $N = 8$, which had $1.25 \times 10^6$ and $0.75 \times
10^6$ configurations, respectively. The configuration energy and center
of mass coordinates were also stored with the cluster coordinates so
that they would not have to be recalculated during subsequent J-walker
simulations whenever a jump was accepted. The required storage space
for these distributions varied from 122 Mb (for $N = 4$) to 580 Mb (for
$N = 24$), and so they were stored in 10 or 50 separate files,
depending on the amount of computer primary memory that was
available.\cite{computers} During a J-walking run, a file was randomly
selected and read into memory. This array was used for sampling jump
attempts for a while before being randomly replaced by another file. By
making the files as large as possible within the memory constraints,
disk access was reduced during the J-walking simulations, greatly
improving their efficiency. Because the configurations generated during
Metropolis walks are highly correlated, they were stored in a periodic
fashion.  Initially, almost all distributions were generated by storing
configurations every 100 passes, as can be seen in
Table~\ref{Tbl:J-walker1} for the smaller clusters. As the cluster size
was increased, generating such distributions became very time
consuming. Not only did it take longer to generate a distribution for a
larger cluster, but more distributions were required to span the entire
temperature range. Thus, for the larger clusters, only the widest
distributions containing the greatest variety of configurations were
sampled every 100 passes. Narrower distributions were sampled every 50
passes, and very narrow distributions for low temperatures
corresponding to the solid region were sampled every 10 passes.
Fig.~\ref{Fig:histogram} shows potential energy histograms for $N = 11$
and $N = 19$ clusters. The smaller cluster has a potential energy range
about half that of the larger cluster, and so only five J-walker
distributions were needed to span the entire temperature range from the
solid region to dissociation, compared to the ten distributions
required for the larger cluster. The spacing of the distributions
requires some judgment. On the one hand, they should be spaced as far
apart as possible to minimize the computational time needed to generate
them. On the other hand, they should not be spaced so far apart that
the number jumps accepted becomes too low and systematic errors arising
from quasi-ergodicity corrupt the distributions. As can be seen in the
plots, substantial overlap between adjacent distributions was required
for reasonable jump acceptances (these are also listed in
Tables~\ref{Tbl:J-walker1} and \ref{Tbl:J-walker2}).

Correlations in the J-walker distributions were further reduced by
writing the distribution files in a parallel fashion. All output
distribution files were opened at the start of the program and
configurations were written to each in turn, rather than writing to
each file sequentially, one at a time. Thus, each distribution file
contained configurations sampled from the entire run. Unfortunately,
this resulted in highly fragmented files, which greatly increased the
time required to later read a distribution into memory, since the files
were all physically interleaved on the disk drive. This problem was
easily solved, however, by simply copying the distribution files to a
different directory, where they were written to the disk sequentially.

Since J-walking is still new, and experience with it is still being
obtained, standard Metropolis simulations were also run for each
cluster. These provided a check of the J-walking results for those
temperature regions where quasi-ergodicity in the Metropolis runs was
not a problem, as well as revealing trends in the systematic errors
arising from quasi-ergodicity in the Metropolis runs as functions of
cluster size. Temperature scans were also generated using Ne
parameters, with the temperature mesh size $\Delta T = 0.02$~K.  For
each temperature, simulations consisted of $10^5$ warmup passes,
followed by $10^7$ passes with data accumulation (again, 100 walks of
$10^5$ passes each). The scans were started at $T = 0.2$ K from the
global minimum configuration,\cite{Northby,Hoare-Pal} and continued
past the cluster dissociation peak. The final configuration for each
temperature was used as the initial configuration for the next
temperature. For all of the clusters examined, the Metropolis and
J-walking results agreed qualitatively throughout the entire
temperature ranges, and agreed quantitatively throughout most of the
temperature ranges, with discrepancies occurring mostly in the
transition regions affecting the peak heights, especially for the
larger sizes; the discrepancies seen in Fig.~\ref{Fig:Ar23} for
Ar$_{23}$ were the largest obtained in this study.

One of the few cluster properties not amenable to calculation using
J-walking is the relative rms bond length fluctuation,
\begin{equation}
    \delta = \frac{2}{N(N-1)}\sum_{i < j}
        \frac{\left(\langle r^2_{ij}\rangle
        - \langle r_{ij}\rangle^2\right)^{1/2}}
        {\langle r_{ij}\rangle}.
\end{equation}
The averages are taken over an entire walk, and so are dependent on the
walk length. With J-walking, a ``walk'' is really a collection of short
excursions of a few steps duration beginning from the different points
in configuration space that are accessed whenever a jump attempt is
accepted, and so there is no continuous evolution of $i$-$j$ bonds over
the entire length of the walk.

Rms bond length fluctuations have been calculated in many previous
molecular dynamics and Metropolis Monte Carlo simulations and have been
used as indicators of solid-liquid phase
changes.\cite{coexist,WB,BJB,EK,DJB} The fluctuations typically undergo
a gradual, nearly linear increase as the temperature is increased in
the solid region, then rise sharply in the transition region as the
clusters begin to acquire enough energy to overcome the potential
barriers to rearrangement, and then level off somewhat in the liquid
region. For magic number clusters such as $N = 13$, the rise can be
very abrupt, while for other clusters, it is more gradual.\cite{BJB}
Like the heat capacity, $\delta(T)$ is very difficult to calculate
accurately in the transition region because of its sensitivity to
quasi-ergodicity. This is evident in Fig.~\ref{Fig:rms}, which shows
averages of $\delta(T)$ for $N = 13$ obtained from 10 Metropolis walks
of $10^4$, $10^5$ and $10^6$ passes in length; the lack of convergence
in the transition region is clear.  The systematic error throughout the
transition region results in the $\delta(T)$ curves for $10^4$ and
$10^5$ passes occurring  too high in the temperature region spanned by
the heat capacity peak, whereas the curve for $10^6$ passes rises sharply
just as the heat capacity curve begins its rise. The curve for $10^5$
passes is similar to the curve reported in Ref.~\onlinecite{DJB}, which
was obtained from Metropolis walks of a similar length. This behavior
can also be understood in terms of the potential energy hypersurface
governing the random walks. For $N = 13$, the deep potential wells
associated with various stable isomers are separated by large
barriers.  Temperatures near the point $T^*_\delta$ where $\delta(T)$
rises sharply correspond to energy distributions having a small
fraction of energies in their tails that are on the threshold of
overcoming the barriers to isomerization. Thus a walk length of $10^6$
passes that resulted in even a few barrier crossings would be averaging
the different fluctuations for different isomers, resulting in much
larger values than those obtained from walks of $10^5$ or $10^4$ passes
where the walker remained trapped in a single isomeric form for the
duration of the walk.

This great sensitivity of the $\delta(T)$ curve on the walk length
brings into question then the appropriateness of applying to clusters
the Lindemann criterion of liquid behavior occurring for $\delta >
0.1$, which has its basis from the study of bulk systems. For clusters,
a value of $\delta \approx 0.1$ obtained from sufficiently long walks
actually corresponds to the threshold of isomerization where the
cluster spends most of its time in some solid-like form with occasional
isomerizations to another solid-like form, rather than to liquid
behavior where the cluster atoms readily undergo diffusion. The region
following the sharp rise where the curve levels off is more
representative of liquid behavior in clusters. This point has been
previously mentioned in accounting for the different energies observed
in molecular dynamics simulations between the onset of large bond
length fluctuations and the onset of diffusion obtained from power
spectra.\cite{BJB} $T^*_\delta$ is also much lower than the value
obtained from molecular dynamics simulations reported in
Ref.~\onlinecite{BJB}, although that value may also be too high
considering the great sensitivity of $\delta$ on the simulation length
found in the Monte Carlo simulations.  Care must be taken when
comparing $\delta$ values obtained from molecular dynamics simulations
with those obtained from Monte Carlo simulations, though. Average rms
bond length fluctuations can depend greatly on the ensemble used, and
discrepancies between results obtained for Ar$_{13}$ in the
microcanonical ensemble with molecular dynamics and those obtained in
the canonical ensemble with Monte Carlo have been noted
previously.\cite{DJB} Despite these problems, $\delta(T)$ curves can
still be very useful for providing insight into the solid-liquid
transition region since their behavior also has a sensitive dependence
on the cluster size.

Rms bond length fluctuations were not originally calculated with the
Metropolis simulations described above, but in order to obtain more
insight into some discrepancies in the transition region between the
J-walking results and results obtained from molecular dynamics
calculations reported in Ref.~\onlinecite{BJB} that became apparent
after the original Metropolis simulations had been run, temperature
scans of $\delta$ were subsequently obtained from additional Metropolis
simulations. These were calculated at each temperature as averages
taken over 10 walks, each having a length of $10^6$ passes.

\section{Results}       \label{Sec:results}
Curves of the heat capacity per atom as functions of temperature for
clusters ranging in size from $N = 4$ to 12 and 13 to 24 are shown if
Figs.~\ref{Fig:Cv1} and \ref{Fig:Cv2}, respectively. These were
obtained from the J-walking studies. As is evident in these figures,
the curves differ markedly and nonmonotonically with cluster size.
There are three kinds of curves. For clusters of size $N = 4$, 5 and 8,
the heat capacity rises slowly from its classical equipartition value
at $T^* = 0$ until the temperature approaches the cluster dissociation
region, where it then rises sharply. There is no peak in the
solid-liquid region, although a very weak shoulder is barely
perceptible. Curves for clusters of size $N = 7$, and 15--17 also have
no peak in the solid-liquid transition region, but they do show a
distinct shoulder characterized by two inflection points, one at a
lower temperature $T^*_L$, about halfway up the shoulder, and the other
at a higher temperature $T^*_H$, at the top of the shoulder. All the
other clusters studied have discernible heat capacity peaks in this
region that can be characterized by an inflection point on the low
temperature side of the peak (at $T^*_L$), the peak maximum itself (at
$T^*_P$), and a local minimum on the high temperature side of the peak
(at $T_H^*$) between the peak and the dissociation region peak. These
temperatures are indicated in Figs.~\ref{Fig:Cv1} and \ref{Fig:Cv2} by
the large ticks on the temperature axes; their values, together with
their corresponding heat capacities, are also listed in
Table~\ref{Tbl:J-walker3}. The values for the peak and local minimum in
each case were obtained by smoothing the J-walker data (note, however,
the curves shown in Figs.~\ref{Fig:Cv1} and \ref{Fig:Cv2} are the raw,
unsmoothed J-walker data), interpolating the smoothed data to obtain a
finer mesh size, and then searching for the maximum and minimum,
respectively. The inflection points were obtained by using finite
differences in the smoothed, interpolated data to generate derivatives,
and then searching the resulting derivative curves for the maximum and
minimum.  Because there was some variation in the values obtained from
various fits generated with different smoothing parameters, several
fits were done for each cluster; the values reported are the averages
obtained from the fits, and the uncertainty estimates are the standard
deviations.\cite{Sav_Gol}

For those clusters having peaks in the solid-liquid region, the
transition temperature can be associated with the peak maximum, and the
transition range with the peak width,\cite{BBDJ} while for those
clusters having shoulders in this region, the transition temperature
can be estimated as the inflection point at the top of the shoulder.
Although the transition temperature thus defined can be obtained easily
for each curve, defining the transition range is more ambiguous. While
the inflection points and local minima that characterize each peak may
be unambiguous, their relation to the width of the transition region is
less clear. In their original J-walker study of classical systems,
Frantz, Freeman and Doll\cite{J-walker} observed that the onset of the
transition region in Ar$_{13}$ could be characterized by a sudden,
sharp rise in the heat capacity standard deviation obtained from
Metropolis Monte Carlo simulations.  For very low temperatures
corresponding to the Ar$_{13}$ solid region, the standard deviation was
very low, and increased slowly with increasing temperature until about
$T^* = 0.20$, where it quickly rose about tenfold, reaching a maximum
at $T^* = 0.25$, and then decreased to a minimum at $T^* = 0.34$ before
rising again in the dissociation region.  If the temperature of the
sharp rise in the heat capacity standard deviation, $T^*_\sigma$, was
taken to signify the beginning of the transition region, and the latter
temperature was taken to signify the end of the transition region, then
the resulting temperature range was found to coincide very well with
the Ar$_{13}$ transition region obtained from molecular dynamics
calculations by Berry and coworkers.\cite{coexist} Moreover, this
behavior of the heat capacity standard deviations was consistent with
their interpretation of the transition region being a coexistence
region marked by sharp, but unequal, melting and freezing temperatures,
$T_m$ and $T_f$; below $T_f$, only solid-like isomers exist, and above
$T_m$, only liquid-like forms exist, but in between in the coexistence
region, both forms dynamically coexist.  Thus, the fluctuations in the
heat capacity were very small for low temperatures since the clusters
were locked into their minimum energy configurations, undergoing small,
mostly harmonic oscillations.  At the freezing temperature, the motion
was highly anharmonic as the clusters began to have sufficient energy
to overcome the large barriers in configuration space separating the
deep wells associated with the different isomers. This hindered
isomerization led to the sharp increase in the heat capacity
fluctuations since the random walks were mostly confined to individual
wells, but occasionally escaped to other wells. The fluctuations
decreased again at the melting temperature because the energy was high
enough for unhindered isomerization to take place; the random walker
had free access to all of configuration space. To better see the
connection between the solid-liquid transition region and the heat
capacity fluctuations, I have also included in Figs.~\ref{Fig:Cv1} and
\ref{Fig:Cv2} standard deviation curves for the heat capacities
obtained from the Metropolis Monte Carlo walks.  These curves are too
noisy to quantitatively define a solid-liquid transition range, but
their general features provide insight into this region. The arrows in
Figs.~\ref{Fig:Cv1} and \ref{Fig:Cv2} indicate the temperatures
$T^*_\sigma$. By taking $T^*_\sigma$ to signify the beginning of the
transition region and $T^*_H$ to mark the end, a definition of the
transition range based solely on heat capacity behavior can be
made,
\begin{equation}
    \Delta T^* = T^*_H - T^*_\sigma.            \label{trans_range}
\end{equation}

The dependence on cluster size for $T^*_\sigma$, as well as for the
heat capacity peak characteristic temperatures, $T^*_L$, $T^*_P$ and
$T^*_H$, is illustrated in Fig.~\ref{Fig:magic}. Also shown is the
dependence on size for their corresponding heat capacities. Because
much of the behavior of cluster heat capacities is influenced by the
structures of the lowest energy isomers, Fig.~\ref{Fig:magic} also
shows the size dependence on the differences in the minimum potential
energy configurations between successive clusters, or the binding
energy differences,\cite{Northby}
\begin{equation}
    \Delta E_b(N) = -\Delta V_{\rm min}(N)
        = -[V_{\rm min}(N) - V_{\rm min}(N-1)]. \label{Vmin}
\end{equation}
The binding energy differences show magic number behavior for $N = 7$,
13, 19 and 23, as evidenced by these sizes being local maxima.

Berry and coworkers have also investigated many of the clusters
included in this study. In one particular molecular dynamics study of
coexistence behavior in rare gas clusters, they reported results for
clusters of size $N = 7$, 8, 9, 11, 13, 14, 15, 17, 19, 20, 22, 26 and
33.\cite{BJB} By checking for the onset and disappearance of a bimodal
distribution in the short-time averaged kinetic energies, they found
coexistence ranges for $N = 7$, 9, 11, 13, 15 and 19.  They also
reported temperatures $T^*_{\delta,MD}$ corresponding to long-time
averaged kinetic energy values at which the rms bond length
fluctuations $\delta$ rose sharply.  For comparison, I have
superimposed these values in Figs.~\ref{Fig:Cv1} and \ref{Fig:Cv2}, as
well as the coexistence ranges $\Delta T^*_c = T^*_m - T^*_f$ for those
clusters where the authors found coexistence behavior. An interesting
feature of the molecular dynamics rms bond length fluctuations for
those clusters showing coexistence behavior is evident in
Figs.~\ref{Fig:Cv1} and \ref{Fig:Cv2}. For the magic number sizes $N =
7$, 13 and 19, $T^*_\delta$ was near the middle of the coexistence
range, nearly as high as the peak temperatures $T^*_P$, while for the
other clusters showing coexistence behavior, $N = 9$, 11, and 15, it
was near the beginning of the range.

\subsection{Pre-icosahedral: $4 \leq N \leq 12$}
The results of the three studies (the J-walking heat capacity curves,
the Metropolis Monte Carlo heat capacity fluctuation curves, and the
molecular dynamics coexistence ranges) are mostly consistent for the
smaller clusters with sizes less than $N = 13$. For $N = 4$ and 5,
there is no abrupt change in either the heat capacity or the heat
capacity fluctuations until the beginning of dissociation region, where
both rise sharply. $N = 6$ is the smallest cluster to show a peak
(albeit a small one) in the heat capacity curve, and the heat capacity
fluctuations show the sharp rise and slow decrease characteristic of a
solid-liquid coexistence region. $N = 7$ is the smallest cluster having
pentagonal symmetry in its lowest energy isomer,\cite{Hoare-Pal} and is
the smallest of the magic number sizes (although it is a rather ``weak''
magic number in comparison to $N = 13$, 19 and 23), and so one might
expect a more pronounced heat capacity peak than found for $N = 6$.
However, $N = 7$ has only a shoulder, although the heat capacity
fluctuations are similar to those for $N = 6$, implying a similar
solid-liquid coexistence region.  Ref.~\onlinecite{BJB} also reported a
coexistence range for $N = 7$, and as can be seen in Fig.~\ref{Fig:Cv1},
this range coincides quite well with the heat capacity transition range
defined by Eq.~\ref{trans_range}. The results for $N = 8$ are very
similar to those obtained for $N = 4$ and 5, and are consistent with
those reported in Ref.~\onlinecite{BJB}.  Wales and Berry\cite{WB}
argued that the striking difference in coexistence behavior between $N
= 7$ and $N = 8$ can be explained in terms of the differences in their
potential energy surfaces. For $N = 7$, the global minimum corresponds
to a very compact, stable structure, the pentagonal bipyramid, which
lies much lower in energy than the other low-lying isomers and is
separated from them by large barriers, preventing any easy
isomerization. For $N = 8$, however, the global minimum corresponds to
a pentagonal bipyramid with the eighth atom located on one of its ten
equivalent faces. Degenerate isomers corresponding to moving the eighth
atom to another face are separated by low barriers, as is a second
isomer, which lies only slightly higher in energy. Thus $N = 8$
clusters have easy access to the various potential wells and exhibit
facile rearrangements. Heat capacity fluctuations remain small until
the dissociation region since the random walker experiences no
bottlenecks in configuration space. Likewise for $N = 4$ and 5.
Although their minimum energy configurations (tetrahedron and trigonal
bipyramid, respectively) are compact and stable, their barriers to
interconversion to other isomers are small.

The heat capacity results obtained for $N = 8$ appear to be contrary to
some of the results obtained by Adams and Stratt,\cite{AS} who
performed an instantaneous normal mode analysis on Ar$_7$, Ar$_8$ and
Ar$_{13}$.  They observed transition regions for both Ar$_7$ and Ar$_8$
based on the temperature dependence of the average percentage of
imaginary frequencies, but found no evidence of a coexistence region in
either. They also reported other dynamical similarities between the two
clusters, and interpreted their results to imply that the solid-liquid
transition should not be viewed primarily in terms of jumps between
rigid and fluid structures, but as being a smooth progression through
an ever increasing number of structural isomers that are locally
similar, but globally distinct. The striking difference between the
heat capacity curves for $N = 7$ and 8, as well as between their heat
capacity fluctuations, seems to indicate an inherently different
transition behavior for these two clusters, and the consistency between
these results and the molecular dynamics results lends further support
to the coexistence view. However, not all the properties show such a
large degree of dissimilarity. Self diffusion constants calculated by
Adams and Stratt for Ar$_7$ and Ar$_8$ as functions of temperature were
very similar,\cite{diffusion} and curves of the rms bond length
fluctuations $\delta$ as functions of temperature for these two
clusters are more alike than different. Although the featureless nature
of the $N = 8$ heat capacity curve might imply that this cluster does
not have a solid-liquid transition region, the diffusion constant
behavior and the rms bond length fluctuation behavior suggest
otherwise. The $\delta(T)$ behavior can be seen in
Fig.~\ref{Fig:rms4-9}, which shows curves for $N = 4$--6 and 7--9. Both
the $N = 7$ and 8 curves show the sharp rise in $\delta(T)$ that is
characteristic of a transition. Although the heat capacity curves for
$N = 4$, 5 and 8 are very similar in their lack of significant features
in the solid-liquid transition region, only the $N = 4$ $\delta(T)$
curve has a likewise featureless rise through the transition region,
implying that no solid-liquid transition occurs before the cluster
dissociates.  Surprisingly then, the absence of a peak or shoulder in
the heat capacity curve does not necessarily correspond to the absence
of a transition. One important difference for the $N = 7$ $\delta(T)$
curve is that its sharp rise occurs at a significantly higher
temperature than that of its immediate neighbors (it is almost as high
as that for $N = 9$).  This was also the case for the molecular
dynamics results reported in Refs.~\onlinecite{WB} and
\onlinecite{BJB}.  As will be seen, this is a common feature of the
other magic number clusters, and is another manifestation of their
inherent stability.

The heat capacity curves for $N = 9$ to 12 all exhibit small heat
capacity peaks in the solid-liquid transition region, and their
Metropolis Monte Carlo heat capacity fluctuations show the sharp rise
and slow decrease characteristic of coexistence behavior.
Ref.~\onlinecite{BJB} reported results for $N = 9$ and 11, and
coexistence regions were found for both clusters. Again, the
coexistence ranges obtained from the molecular dynamics calculations
coincide very well with the transition ranges obtained from the heat
capacities. The lowest energy isomers for these clusters represent the
building up from $N = 7$ by the successive addition of atoms to the
five faces of the pentagonal bipyramid core, with the $N = 12$
structure then being one atom short of an icosahedron. This regular
progression is reflected in the heat capacity curves. The $N = 9$ peak
is small and close enough to the dissociation peak that it is barely
more than a shoulder. The $N = 10$ and $N = 11$ peaks are slightly
larger, and remarkably alike. The $N = 12$ peak is considerably larger
than the others.  The transition ranges also show quite a regular
progression, as is evident in Fig.~\ref{Fig:magic}, with the $N = 9$
range being the smallest and occurring at the lowest temperature. The $N
= 10$, 11 and 12 transition ranges are about equal in size, but occur
at increasingly higher temperatures.

\subsection{Icosahedral to Double Icosahedral: $13 \leq N \leq 19$}
The results for the range $N = 13$ to 19 are consistent with the
molecular dynamics results for the magic number endpoints $N = 13$ and
19, but there are some surprising differences for sizes in between.  $N
= 13$ is the prototypical magic number cluster and has been the subject
of many investigations.  Its minimum energy configuration is the
icosahedron, a compact structure that is extremely stable. The $N = 13$
potential energy surface is characterized by very deep wells,
corresponding to the icosahedral isomers, which are separated by large
barriers and lie far below the wells corresponding to the next lowest
energy isomers. Its heat capacity behavior is a reflection of the shape
of this potential energy surface.  The curve has a very large,
pronounced peak in the solid-liquid region (more than double the height
of the preceding $N = 12$ peak), and the fluctuations in the heat
capacities obtained from Metropolis Monte Carlo simulations show the
characteristic features of a coexistence region. The coexistence range
obtained from molecular dynamics calculations reported in
Ref.~\onlinecite{BJB} also coincides very well with the transition
region found in this study.

The J-walking results also agree with the original J-walking results
for Ar$_{13}$ reported in Ref.~\onlinecite{J-walker}, except for a
small discrepancy in the peak height where a value of $\langle C^*_V
\rangle/N = 9.37 \pm 0.04$ was obtained at $T^*_P = 0.285 \pm 0.004$,
compared with $\langle C^*_V \rangle/N = 9.06 \pm 0.01$ obtained at
$T^*_P = 0.2875 \pm 0.0007$ in this study. This investigation was more
thorough than the original one. In that study, computer constraints
limited the J-walker distribution sizes to only $5 \times 10^4$
configurations, and walk lengths for each temperature consisted of
$10^7$ moves. In this study, J-walker distributions were twenty times
larger, and walks consisted of $10^7$ passes, and so were thirteen
times longer. In addition, the first J-walker distribution in the
original investigation was obtained from a long Metropolis Monte Carlo
walk at $T^* = 0.42$. This was a temperature thought to be in the
liquid region (and thus presumed to be free of quasi-ergodicity), but
was actually near the beginning of the dissociation region where
quasi-ergodicity again begins to become problematic. Since subsequent
lower temperature J-walker distributions were generated in sequence
from this distribution, any systematic errors in the distribution could
have propagated to the other distributions and thus affected the
J-walking results. In this study, the initial J-walker distribution was
generated at $T^* = 0.618$, a temperature on the high temperature side
of the dissociation region heat capacity peak where the heat capacity
fluctuations had dropped substantially, thus reducing systematic errors
in the lower temperature distributions.  Despite these problems, the
earlier J-walking study still gave a value for the heat capacity peak
that was within 4\% of the value obtained in this study for not much
more computational overhead than the similar Metropolis Monte Carlo
simulation also reported in Ref.~\onlinecite{J-walker}, which had an
error of 19\% for the peak height.

Further support for the accuracy of the results reported here can be
been found in a recent study by Tsai and Jordan,\cite{Tsai-Jordan2} who
combined histogram and J-walking methods to calculate heat capacity
curves for Ar$_{13}$. They obtained a peak height of $\langle C^*_V
\rangle/N = 9.01$ at $T^*_P = 0.286$, which is very close to the value
reported here; the size of the difference between my values and
theirs is similar to the small variations they observed for results
obtained using different constraining sphere radii (their results were
obtained using a smaller constraining radius, with $R_c = 10$ \AA, or
2.94$\sigma$, compared to $R_c = 4\sigma$ used in this study).
Finally, the results obtained from the Metropolis runs of the same
length ($10^7$ passes) agree with the J-walker results, with
$\langle C^*_V \rangle/N = 8.97 \pm 0.04$ and $T^*_P = 0.2865 \pm
0.0011$.  This indicates that Metropolis walk lengths of $10^7$ passes
are sufficiently long to overcome quasi-ergodicity for $N = 13$,
consistent with the findings of Tsai and Jordan, who reported that walk
lengths of $6 \times 10^6$ passes were sufficiently long.

The results obtained for $N = 14$ to 18 are surprising, and are
considerably different from the results obtained for $N = 8$ to 12.
This was unexpected since the two ranges are analogous in many
respects. Both are bounded by magic number sizes ($N = 7$ and 13, and
$N = 13$ and 19, respectively) and have a similar ``growth'' sequence
in their minimum energy configurations. This can be seen in
Fig.~\ref{Fig:clusters}, which shows the minimum energy configurations
for $N = 8$ to 19.\cite{min_config}  Where the $N = 8$ lowest energy
isomer consists of a lone atom located on one of the ten faces of the
$N = 7$ pentagonal bipyramid core, and where the succeeding lowest
energy isomers can be formally obtained by the addition of atoms in a
fivefold symmetric manner, with the sixth capping the ring to form the
$N = 13$ icosahedron, the $N = 14$ lowest energy isomer likewise
consists of a lone atom located on one of the twenty equivalent
icosahedral faces, and the succeeding lowest energy isomers (except $N =
17$) can be likewise formally obtained by the addition of atoms in a
fivefold symmetric manner, with the sixth capping the ring to form the
$N = 19$ double icosahedron (the expected $N = 17$ configuration
obtained in this manner is only slightly higher in energy than the
actual minimum energy isomer shown, with an energy $V^* = -61.3071$,
compared to $V^*_{\rm min} = -61.3180$). The striking similarity in the
binding energy differences for these two ranges can be seen in the plot
of $\Delta E_b(N)$ shown in Fig.~\ref{Fig:magic} --- the sequence for
$N = 8$ to 13 is nearly identical to the sequence for $N = 14$ to 19.
Molecular dynamics calculations reported in Ref.~\onlinecite{BJB} also
found analogous behavior. Both Ar$_7$ and Ar$_{13}$ had relatively
large coexistence ranges, while neither Ar$_8$ nor Ar$_{14}$ showed any
coexistence behavior, and both Ar$_9$ and Ar$_{15}$ showed coexistence
behavior, but had narrower ranges than Ar$_7$ and Ar$_{13}$.  Wales and
Berry\cite{WB} ran extensive molecular dynamics simulations combined
with quench studies for Ar$_7$, Ar$_8$, Ar$_{13}$ and Ar$_{14}$ that
confirmed the earlier results.  Yet, the heat capacity curves obtained
with J-walking depicted in Figs.~\ref{Fig:Cv1} and \ref{Fig:Cv2} are
not at all consistent with these results. Where the $N = 8$ heat
capacity curve is featureless in the solid-transition region, the $N =
14$ curve shows a clear peak. The peak is much diminished relative to
the $N = 13$ peak, but is nonetheless quite substantial; it is larger
than the $N = 12$ peak, which is the largest peak for $N < 13$. The
fluctuations in the Metropolis heat capacities are consistent with the
J-walking results, showing only a moderately steep rise followed by a
slow decrease, suggesting some coexistence behavior.  The results for
$N = 15$ are even more surprising in comparison. Ref.~\onlinecite{BJB}
reported a coexistence region for Ar$_{15}$, analogous to Ar$_9$, but
the J-walking results reveal only a shoulder in the heat capacity curve
over this region.  While this does not rule out a coexistence region
(as evidenced by the shoulder in the heat capacity curve for $N = 7$,
which does show coexistence behavior), it does, then, make the absence
of the coexistence region in the $N = 14$ molecular dynamics
simulations look even more puzzling by comparison. The behavior of the
Metropolis heat capacity fluctuations for $N = 15$ is also very
different; the typical sharp rise signifying the onset of the
coexistence region has been replaced by a more gradual increase that
rises slowly over the domain spanned by the shoulder before beginning
the usual rise in the dissociation region. The inflection points
characterizing the heat capacity curve occur well after the molecular
dynamics coexistence region, and the rise in the fluctuations occurs
well before, resulting then in a very much larger transition region.

Rms bond length fluctuation curves for $N = 13$, 14 and 15 are shown in
Fig.~\ref{Fig:rms13_19}. As with $N = 7$, magic number behavior is
manifested in $N = 13$ by its much higher temperature $T^*_\delta$
signifying the sharp rise in $\delta(T)$. Further similarities between
$N = 14$ and $N = 15$ are evident in the $\delta(T)$ curves. Except for
the unusual smaller rise in $\delta(T)$ beginning at $T^* = 0.064$ for
$N = 14$, the two curves differ very little. This smaller rise is
likely due to the facile movement of the extra atom for $N = 14$ on the
icosahedral core. The average potential energy corresponding to this
temperature is $V^* = -46.60$, which is comparable to the saddle point
energy of $-47.06$ representing the edge-bridging isomerization that
moves the fourteenth atom from one icosahedral face to a neighboring
one. It is also comparable with two other saddle point energies of
$-45.95$ and $-45.84$, which also represent isomerizations to
degenerate capped icosahedrons.\cite{WB}

$N = 16$ continues the trend from $N = 15$, with the heat capacity
curve reduced to an even smaller shoulder, and with the Metropolis heat
capacity fluctuations remarkably uniform throughout the solid-liquid
region. The results for $N = 17$, 18 and 19 mirror the results for $N =
13$, 14 and 15, with $N = 17$ having a shoulder in its heat capacity
curve similar to the $N = 15$ curve, $N = 18$ having a moderate peak
similar to the $N = 14$ peak, and $N = 19$ having pronounced peaks in
both the heat capacity and the Metropolis heat capacity fluctuation
curves, consistent with its magic number status.  The only significant
difference is that the shoulder and peak sizes for $N = 17$, 18 and 19
are smaller than their counterparts for $N = 15$, 14 and 13.
Interestingly, the rms bond length fluctuation curve for $N = 18$ also
mirrors the $N = 14$ curve with a smaller rise beginning at $T^* =
0.106$ and ending at $T^* = 0.168$, where it rises sharply.  Further
evidence of magic number behavior in $N = 19$ is evident in the
$\delta(T)$ curves shown in Fig.~\ref{Fig:rms13_19}, where once again
$T^*_\delta$ can be seen to occur at a much higher temperature than
that of the next two larger clusters. The molecular dynamics
coexistence range for $N = 19$ reported in Ref.~\onlinecite{BJB} also
coincides quite well with the transition range I obtained, although it
is slightly contracted.

\subsection{Beyond Double Icosahedral: $20 \leq N \leq 24$}
The results for $N = 20$ to 24 also differ substantially from those of
$N = 14$ to 18. This is not unexpected since the minimum energy
configurations for these clusters show a considerably different
structural sequence, with additional atoms occupying sites between the
three equatorial 5-atom rings of the $N = 19$ double icosahedron.
Fig.~\ref{Fig:clusters2} shows the minimum energy configurations for $N
= 20$ to 24. For $N = 23$, the additional atom caps an interpenetrating
icosahedral substructure, resulting in a magic number configuration.
Magic number behavior can be seen in Figs.~\ref{Fig:Cv2} and
\ref{Fig:magic} by both the large heat capacity peak and in the binding
energy difference maximum. The heat capacity peak is about the same
size as the previous magic number cluster, $N = 19$. The Metropolis
heat capacity fluctuations also show the sharp rise and larger peak
indicative of a coexistence region.

Unlike $N = 14$, whose heat capacity peak is much smaller than its
preceding magic number $N = 13$ peak, the peaks for $N = 20$ and 24 are
only slightly smaller than their preceding magic number peaks. The
major differences between $N = 19$ and 20, as well as between $N = 23$
and 24, lie at the beginning of the transition regions. While the heat
capacity fluctuations for $N = 19$ and $N = 23$ show the sharp rise and
subsequent slow decline characteristic of a coexistence region, the
fluctuations for $N = 20$ and $N = 24$ show a more gradual rise to a
smaller maximum, followed by a slow decline. This was also the case for
$N = 21$ and 22, and so determining the onset of the transition region
for each of these clusters was ambiguous. As with $N = 14$, there was no
evidence from molecular dynamics calculations of a coexistence region
reported for $N = 20$ in Ref.~\onlinecite{BJB}, which is surprising since
its heat capacity peak is nearly as large as that for $N = 19$ and its
heat capacity fluctuations are also quite pronounced.  The heat capacity
curves and Metropolis heat capacity fluctuations are remarkably alike for
$N = 21$ and 22, reminiscent of the similarity between $N = 10$ and 11.
Both $N = 21$ and 22 also have smaller peaks than the other clusters in
this range. Molecular dynamics results for $N = 22$ were also reported in
Ref.~\onlinecite{BJB} and no evidence of a coexistence region was found.

\section{Discussion}                    \label{Sec:discuss}
Although the heat capacity-temperature curves do not provide the
definitive measure of the cluster solid-liquid transition region, they
do serve as yet another useful probe into this region, providing
additional data of surprising diversity and complexity that needs to be
examined in terms of the various theories that have been proposed.
Many of the results reported here are consistent with the view that for
some clusters (especially for the magic number sizes), the solid-liquid
transition region corresponds to a coexistence region with sharply
defined, but separate, melting and freezing temperatures.  Just how
sharp these transition temperatures really are remains an open
question, but for some clusters, properties obtained from Metropolis
simulations such as the heat capacity fluctuations and rms bond length
fluctuations do exhibit threshold-temperature behavior, with abrupt
changes occurring over narrow temperature ranges. For many of the
clusters where molecular dynamics simulations found evidence of
coexistence behavior, the Metropolis heat capacity fluctuations were
consistent, and the transition ranges obtained from Metropolis heat
capacities coincide quite well with the coexistence ranges.  But there
are also some puzzling discrepancies, such as with $N = 14$ and 15.
Molecular dynamics simulations found evidence of coexistence behavior
for $N = 15$, which has neither a heat capacity peak nor an appreciable
change in the heat capacity fluctuations throughout the solid-liquid
transition region, but found no coexistence behavior for $N = 14$,
which does have a heat capacity peak and an increase in the
fluctuations. On the other hand, the view of the solid-liquid
transition being a series of isomerizations through an ever increasing
number of structures also appears consistent with my findings for some
of the clusters studied, but it seems hard pressed to account for the
entire range of behavior seen in the heat capacity curves and rms
fluctuations. In addition, some supporting evidence is contrary to the
heat capacity behavior. For example, the instantaneous normal mode
analysis of $N = 7$ and 8 indicated similar behavior in the transition
region, but the heat capacity studies showed them  behaving very
differently. Similarly, the instantaneous normal mode analysis of $N =
13$ indicated no abrupt changes for some dynamical properties in the
transition region, where the heat capacity, its fluctuations, and the
rms bond length fluctuations all show abrupt changes.

It is also clear (for these smaller clusters, at least) that heat
capacity peaks by themselves are incomplete measures of the
solid-liquid transition region. For example, both $N = 19$ and $N = 20$
have similarly sized heat capacity peaks, but $N = 19$ also shows magic
number behavior in its binding energy and rms bond length fluctuations,
and shows coexistence behavior according to molecular dynamics
simulations, while $N = 20$ exhibits none of these.  Even the absence
of a heat capacity peak in the solid-liquid region is no indication of
the absence of a transition there. The curves for $N = 4$, 5 and 8 are
all featureless in this region, but rms bond length fluctuation curves
for $N = 5$ and 8 show evidence of a solid-liquid transition. $N = 7$
is a magic number cluster, but its heat capacity curve has no peak,
even though its non-magic number predecessor $N = 6$ does have a heat
capacity peak.

Additional insight into the solid-liquid transition region can be obtained
by comparing the heat capacity magic number behavior with that of other
properties.  Fig.~\ref{Fig:magic} shows that the heat capacity magic
numbers are the same as the $\Delta E_b(N)$ magic numbers over the range
$4 \leq N \leq 24$, but there are interesting differences between the two
sequences.  The magic numbers for the $\Delta E_b(N)$ sequence are $N =
7$, 13, 19 and 23, as evidenced by these sizes having the largest values
(except for $N = 7$, which has the largest value for $N < 9$). In each
case, $\Delta E_b(N)$ substantially decreases for $N$ immediately
following, and then rises. The magic numbers for the heat capacity are
likewise $N = 7$, 13, 19 and 23, which are the local maxima in the heat
capacity peak sequence.  However, with the exception of $N = 8$, which
has a featureless transition region, the peaks for $N = 14$, 20 and 24 do
not show a corresponding dramatic decline followed by an increase.
Instead, there is a gradual decline over several sizes.  $N = 14$, which
has a small peak, is followed by three clusters having smaller shoulders,
while $N = 20$ has nearly as large a peak as the $N = 19$ peak, and $N =
24$ has a peak only slightly smaller than the $N = 23$ peak. Furthermore,
the heat capacity peak sequence for $N = 8$ to 13 does not show the
remarkable similarity with the $N = 14$ to 19 sequence that is evident in
the $\Delta E_b(N)$ values. Thus, it appears a cluster's minimum energy
structure has a dominant influence on the  heat capacity only for the
magic number sizes.

The impact of the minimum energy structure on cluster energetics for
magic number sizes is also evident in Fig.~\ref{Fig:potential}, which
plots for all the clusters studied the potential energies per atom as
functions of temperature.  The curves have a roughly uniform spacing
for temperatures corresponding to the liquid region, while for
temperatures in the solid region, the curves show the nonuniform
spacing expected for the very different geometrical structures
associated with each minimum energy configuration, with the enhanced
stability of the magic number clusters $N = 13$, 19 and 23 (and to a
lesser extent, $N = 7$) clearly evident. Also included in the plot are
the $T^*_\sigma$ values and the heat capacity peak temperatures,
$T^*_L$, $T^*_P$ and $T^*_H$, which characterize the transition region.
The nonuniform spacing of the solid region and the enhanced stability of
the magic number structures persist through much of the transition
region, as does the uniform spacing of the liquid region. This indicates
that both solid-like and liquid-like forms are indeed found in the region
spanned by the heat capacity peak.  The persistence of solid-like forms
into the transition region and the enhanced structural stability of the
magic number sizes can be seen even more dramatically in
Fig.~\ref{Fig:DV}, where the $\Delta E_b(N) = -\Delta V^*(N)$ curve for
zero temperature shown in Fig.~\ref{Fig:magic} has been generalized to
nonzero temperatures to form the surface $-\Delta \langle
V^*(N,T)\rangle = -[\langle V^*(N,T)\rangle - \langle
V^*(N-1,T)\rangle]$, with the temperature range spanning the
solid-liquid transition region up to the beginning of the dissociation
region of the larger clusters. The icosahedral based magic number sizes
$N = 13$, 19 and 23 are especially noticeable --- the $-\Delta \langle
V^*\rangle$ peak values differ very little from their zero temperature
values well into the transition region, until their abatement over a
relatively narrow temperature range at the heat capacity peak
temperature. This temperature range is much smaller than the transition
range defined in Eq.~\ref{trans_range}, beginning at about $T^*_L$
rather than at $T^*_\sigma$ and ending well before $T^*_H$.  Although
this seems to indicate more the abrupt change from solid to liquid
rather than the coexistence of solid and liquid forms over an extended
range, it must be remembered that these energy differences are between
average energies, which can be quite insensitive to the makeup of their
underlying distributions.

The average potential energy and configurational heat capacity at a
given temperature are of course just the first two moments of the
potential energy distribution for that temperature, and so they
represent only a minimal description of the distribution. Further
insight into cluster energetics can be obtained by examining the
distributions themselves. With J-walking, accurate potential energy
distributions can be obtained during a simulation at a given
temperature by binning the potential energies during the walk and
forming a histogram. For example, the distributions shown in
Fig.~\ref{Fig:histogram} were obtained this way. By projecting several
histograms over a temperature range, a distribution surface can be
formed as a function of the potential energy and temperature.
Fig.~\ref{Fig:surf_dis} shows examples of such surfaces. These are for
the range $N = 13$ to 18, which contains representative samples of the
different types of heat capacity behavior encountered (a strong peak at
$N = 13$ declining to a weak shoulder at $N = 16$ and then rising again
to a strong peak at $N = 19$). Narrow distributions occur at the
temperature extremes corresponding to the solid and dissociation
regions; saddle points correspond to the widest distributions and thus
represent heat capacity peaks. Not surprisingly, the change in the
distributions in the solid-liquid transition region is much more abrupt
for $N = 13$ than for the other clusters. Not only do the distributions
for the solid region widen appreciably and then narrow suddenly for the
liquid region to form the characteristic hump on the surface, but the
distributions for the two regions are partially offset, indicating a
bimodal distribution. This is evident in Fig.~\ref{Fig:magic_dis}, which
shows the potential energy distributions for temperatures near the heat
capacity peak temperature $T^*_P$. The bimodal distribution indicates
well separated energies that can be associated with distinct solid-like
and liquid-like forms and thus are further evidence of coexistence
behavior in $N = 13$. Unfortunately, $N = 13$ is the only cluster in
the range studied that clearly has a bimodal distribution in the
solid-liquid transition region. As can be seen in
Fig.~\ref{Fig:magic_dis}, even the other magic number clusters $N = 19$
and $N = 23$, which have similar distribution surfaces with a similar
widening at the heat capacity peak temperature, do not show clear
bimodality there. This does not rule out coexistence behavior, though,
since the two distributions might overlap enough to show only a single
maximum, but it does indicate that the different distributions are not
widely spaced apart, like they are in $N = 13$.

The distribution surfaces for $N = 15$, 16 and 17 show a very smooth,
gradual evolution from the sharp, narrow distributions in the solid
region to the broader distributions in the liquid region. This behavior
is consistent with the view of the solid-liquid transition being a
progression through an increasingly large number of structural isomers.
For these clusters then, there are no threshold changes in the potential
energies being accessed as the temperature increases past the solid
region --- higher energy isomers with energies slightly higher than the
minimum energy isomer are separated by low barriers and thus are easily
accessible. $N = 16$ clusters have several local minima lying close to
the global minimum,\cite{HA} as does $N = 17$,\cite{coexist} which
also has a local minimum differing by less than 0.02\%. The similarity
between these three clusters is also evident in the behavior of their
average potential energy per atom. As can be seen in
Fig.~\ref{Fig:potential}, the spacings between the curves for these
clusters are almost equal and change very little throughout the
temperature range. Again, this is very much different than the behavior
displayed by the magic number clusters.

Further evidence of the coexistence between solid-like and liquid-like
isomers for the magic number clusters in the transition region comes
from the rms bond length fluctuation curves. Comparison of the
$\delta(T)$ curves shown in Fig.~\ref{Fig:rms13_19} with the $-\Delta
\langle V^*(N,T)\rangle$ surface in Fig.~\ref{Fig:DV} reveals that the
temperatures $T^*_\delta$ where the fluctuations for each cluster rise
sharply from their nearly linear increase in the solid region (thus
marking the onset of hindered isomerization) are well below the
temperatures where the $-\Delta \langle V^*(N,T)\rangle$ peaks diminish
quickly, implying that the solid-like forms persist well into the
transition region (although undergoing increasingly frequent
isomerizations). However, the end of the sharp rise and the leveling of
the $\delta(T)$ curves to a value of about 0.3 also occurs in this
region, indicating liquid-like behavior there as well. Note also that the
lack of change in the average potential energy difference between $N =
14$ and $N = 13$ up to a reduced temperature of about 0.25 lends
further support to the conjecture that the initial rise in $\delta(T)$
for $N = 14$ is due to the motion of the lone atom on the icosahedral
core, rather than an isomerization to one of the other low-lying
configurations.

Magic number behavior in $T^*_\delta$ is shown in
Fig.~\ref{Fig:magic_rms}, which plots the $T^*_\delta$ values obtained
from the Metropolis simulations against cluster size. Magic number
behavior is evident for $N = 13$, 19 and 23 (and to a lesser extent for
$N = 7$) by their much higher $T^*_\delta$ values. The $T^*_\delta$
values for all the clusters lie below $T^* = 0.18$, which according to
Figs.~\ref{Fig:potential} and \ref{Fig:DV} places them near the end of
the solid region. Also, as can be seen in Fig.~\ref{Fig:magic_rms},
they generally coincide with the $T^*_\sigma$ values obtained from the
rise in the heat capacity fluctuations, further supporting the claim
that this region marks the beginning of the transition. Given the large
degree of ambiguity in determining either $T^*_\sigma$ or $T^*_\delta$
for some of the larger clusters because of the large levels of noise in
their fluctuations, it is not valid to conclude much more than the
general observation that these values increase slightly with increasing
cluster size, except for magic number sizes, where they occur at much
higher temperatures.

Also included in Fig.~\ref{Fig:magic_rms} are the $T^*_{\delta,MD}$
values obtained from molecular dynamics simulations reported in
Ref.~\onlinecite{BJB}.  Although these values are all significantly
higher than the Metropolis values, their pattern is quite similar, with
the magic number clusters having values that are substantially higher
than those of their immediate neighbors. As noted earlier, the
discrepancies between the two sets of data result from the shorter
trajectory lengths used in the molecular dynamics simulations, and from
the different ensembles associated with each method.\cite{DJB}

Those clusters exhibiting coexistence behavior also have their
coexistence ranges shown in Fig.~\ref{Fig:magic_rms}. Except for $N =
15$, which has a coexistence range, contrary to the heat capacity
behavior, and $N = 20$, which has no coexistence range even though the
heat capacity data suggests it should, the correspondence between the
coexistence ranges and the transition ranges defined by the heat
capacity behavior is quite good. For $N = 7$, 9, 11, 13 and 19, both
studies give consistent evidence of coexistence behavior, and the size
and location of the two ranges are similar, while for $N = 8$, 17 and
22, the lack of evidence of coexistence behavior is likewise
consistent. Unlike most of the molecular dynamics results used as
evidence for the coexistence region, though, the heat capacity results
obtained from the Monte Carlo simulations are equilibrium properties
and thus provide mostly inferential, not direct, evidence of dynamical
coexistence between solid-like and liquid-like forms. The large degree
of consistency between the two does lend strong support to the
hypothesis, but the discrepancies for $N = 15$ and $N = 20$ are
troubling in that they have no obvious resolution. While it appears
that quasi-ergodicity is an even more insidious problem than has been
appreciated in the past, one cannot simply dismiss these discrepancies
as being due to improper sampling in the molecular dynamics simulations
without accounting for the good agreement for the other sizes, which
were simulated in similar manner (also, it is the magic number sizes
that are most prone to quasi-ergodicity problems, and these are the
sizes where agreement is the strongest). The discrepancies might be the
result of the different ensembles used,\cite{Stratt} since some cluster
properties have a strong ensemble dependence. However, comparison of
ensemble sensitive properties such as the rms bond length fluctuations
and diffusion constants that have been calculated in both the
micro-canonical and canonical ensembles reveals more similarities than
differences (the the temperature curves have roughly the same shape and
differ primarily by their displacement in temperature) and again, one
is left with the dilemma of justifying why the agreement between the
Monte Carlo results and the molecular dyanamics results is so poor for
these clusters but so good for the other sizes.

Despite these inconsistencies, the bulk of my results are consistent
with the view of a dynamic coexistence in the transition region and can
be interpreted within that framework. My results are not totally
inconsistent with the viewpoint that emphasizes the smooth progression
of isomerizations between locally similar but globally distinct
configurations occurring throughout the transition region. Clearly,
cluster transitions are a progression of isomerizations --- the issue
is just how smooth these progressions really are. For clusters such as
$N = 15$, 16 and 17 the progressions appear to be very smooth, but for
many other clusters (especially the magic number clusters), the
threshold behavior of many of their properties and their very different
heat capacity behavior implies an abruptness in the progressions that
is more appropriately described then by the coexistence viewpoint.

The nature of cluster phase transitions still remains much unresolved
then, despite the additional information provided by the heat capacity
results. More data is required to resolve the outstanding issues.  This
can be accomplished by extending the systematic survey of cluster
properties beyond $N = 24$, but another avenue also seems appropriate.
Since many cluster properties result from an interplay of particle
dynamics based on coordinated atom moves,  and structural effects based
on particle size and intermolecular forces, the study of heterogeneous
atomic clusters of various size and composition can help determine the
relative importance of each factor. Work is currently underway on a
J-walking study of binary Ne-Ar and Ar-Kr clusters. Ne and Ar differ
greatly in both size and potential, but Ar and Kr have similar sizes
and so differences in their behavior can provide more insight into the
transition region.

\acknowledgments
Support for this research by the Natural Sciences and Engineering
Research Council of Canada (NSERC) is gratefully acknowledged.  I also
thank David L. Freeman for helpful discussions, and Richard M. Stratt
for his comments.

\begin{figure}
\caption{Classical internal energy (at left) and constant volume heat
capacity (at right) as functions of temperature for Ar$_{23}$ clusters;
both are in reduced units, with $U^* = U/\epsilon$ and $C_V^* =
C_V/k_B$. The lower temperature heat capacity peak corresponds to a
solid-liquid transition, while the higher temperature peak corresponds
to a liquid-vapor transition. The solid curves were obtained using
J-walking from external J-walker distributions containing $10^6$
configurations; the jump attempt frequency was $P_J = 0.1$. For each
temperature, $10^5$ warmup passes were run, followed by $10^7$ passes
with data accumulation. The dashed curves were obtained using standard
Metropolis Monte Carlo methods, likewise consisting of $10^5$ warmup
passes followed by $10^7$ passes with data accumulation.  Some
representative single standard deviation error bars have been
included.  Also included in the heat capacity plot are results from
Metropolis runs consisting of $10^5$ warmup passes followed by $10^5$
passes with data accumulation (curve i) and $10^6$ passes with data
accumulation (curve ii). Lack of convergence in the Metropolis
simulations is evident in the solid-liquid transition region.
\label{Fig:Ar23}}
\end{figure}

\begin{figure}
\caption{Normalized potential energy histograms for $N = 11$ and $N =
19$ J-walker distributions. Distribution particulars are given in
Tables~\protect\ref{Tbl:J-walker1} and \protect\ref{Tbl:J-walker2}. As
the cluster size increases, so does the potential energy range,
implying more distributions are required to span the entire range.
\label{Fig:histogram}}
\end{figure}

\begin{figure}
\caption{The dependence on walk length for Metropolis averages of the
rms bond length fluctuations $\delta$ as a function of temperature for
$N = 13$. The values at each temperature are averages obtained from 10
Metropolis Monte Carlo walks having lengths of $10^4$ passes (short
dashes), $10^5$ passes (long dashes) and $10^6$ passes (solid line).
The dotted curve is the heat capacity from Fig.~\protect\ref{Fig:Cv2},
and has been included for comparison.  Slow convergence due to
quasi-ergodicity in the transition region is evident by the
displacement of the curves to lower temperatures with increased walk
length. The insert shows the temperature $T^*_\delta = 0.173$ where the
rms bond length fluctuations begin to rise sharply. The solid triangle
indicates the temperature $T^*_{\delta,{\rm MD}}$ where the rms bond
length fluctuations obtained from molecular dynamics simulations
reported in Ref.~\protect\onlinecite{BJB} were found to rise sharply.
\label{Fig:rms}}
\end{figure}

\begin{figure}
\caption{Heat capacity curves as functions of temperature (upper curves
in each plot, in reduced units) for clusters ranging from $N = 4$ to 12.
All curves where obtained using J-walking from externally stored
J-walker distributions with jumps attempted with a frequency of $P_J =
0.1$. The distribution parameters are listed in
Table~\protect\ref{Tbl:J-walker1}. For each temperature, simulations
consisted of $10^5$ warmup passes followed by $10^7$ passes with data
accumulation. The lower, noisy curves are heat capacity standard
deviations obtained from similar Metropolis Monte Carlo simulations, each
consisting of $10^5$ warmup passes followed by $10^7$ passes with data
accumulation. The temperatures corresponding to a sudden increase in
standard deviation (indicated by the arrows) serve as estimates to the
beginning of the solid-liquid transition regions.  For those clusters
having heat capacity curves with peaks in the solid-liquid transition
region ($N = 6,$ 9--12), the temperatures corresponding to the low
temperature side inflection point, $T_L^*$, the peak, $T_P^*$, and the
high temperature side local minimum, $T_H^*$ are indicated on the
temperature axis by the large ticks. For $N = 7$, which has a shoulder in
its heat capacity curve in this region, the large ticks indicate the
inflection points, $T_L^*$ and $T_H^*$.  These peak parameters are also
listed in Table~\protect\ref{Tbl:J-walker3}.  For those clusters having
molecular dynamics simulations reported in Ref.~\protect\onlinecite{BJB}
($N$ = 7--9 and 11), the solid triangles indicate the temperatures at
which the rms bond length fluctuations $\delta$ were found to rise
sharply, while the horizontal error bars represent the coexistence
temperature ranges, $\Delta T^*_c = T^*_m - T^*_f$, for those clusters
found to have a bimodal form in the distribution of short-time averaged
kinetic energies. \label{Fig:Cv1}} \end{figure}

\begin{figure}
\caption{Same as Fig.~\protect\ref{Fig:Cv1}, but for clusters ranging
in size from $N = 13$ to 24. The J-walker distribution parameters are
listed in Table~\protect\ref{Tbl:J-walker2}.
\label{Fig:Cv2}}
\end{figure}

\begin{figure}
\caption{Magic number effects in cluster heat capacities. The top plot
shows the solid-liquid transition region heat capacity peak
characteristic temperatures as functions of cluster size $N$. For those
clusters having distinct peaks, the values $T_L^*$, $T_P^*$ and $T_H^*$
are indicated, while for those clusters having heat capacity shoulders
in this region, $T_P^*$ is taken to be $T_H^*$.  These points are also
indicated in Figs.~\protect\ref{Fig:Cv1} and \protect\ref{Fig:Cv2} by
the large ticks on the temperature axes. No solid-liquid transition
region heat capacity peaks or shoulders were found for $N = 4$, 5 and
8. The lower $T_\sigma^*$ curve indicates the dependence on cluster
size for the beginning of the transition region (also indicated by the
arrows in Figs.~\protect\ref{Fig:Cv1} and \protect\ref{Fig:Cv2}).  The
solid-liquid transition ranges can be defined as the difference between
the top and the bottom curves, $\Delta T^* = T^*_H - T^*_\sigma$. The
middle plot shows the size dependence on the corresponding heat
capacities.  The bottom plot was inspired by
Ref.~\protect\onlinecite{Northby} and has been included for comparison.
It plots the binding energy differences between successive clusters, as
a function of size, $\Delta E_b(N) = -\Delta V_{\rm min}(N) = -[V_{\rm
min}(N) - V_{\rm min}(N-1)]$.  Magic number sizes based on zero
temperature structural stability ($N = 13$, 19 and 23, and to a lesser
extent, $N = 7$) have a clear correlation with those evident in the
heat capacity peaks.
\label{Fig:magic}}
\end{figure}

\begin{figure}
\caption{Rms bond length fluctuations as functions of temperature for $N
= 4$--9. The values at each temperature are averages obtained from 10
Metropolis Monte Carlo walks having lengths of $10^6$ passes. Some
representative single standard deviation error bars have been included.
Only the $N = 4$ curve shows no abrupt change in the solid-liquid
region before rising sharply in the dissociation region.  The $N = 7$
curve shows magic number behavior that reflects the enhanced stability
of the cluster's lowest energy configuration.  Its sharp rise occurs at
a much higher temperature than that for either $N = 6$ or $N = 8$.
\label{Fig:rms4-9}}
\end{figure}

\begin{figure}
\caption{Pentagonal ``auf-bau'' in the minimum energy configurations
for Lennard-Jones clusters. The configurations shown are the lowest
energy isomers obtained from the J-walking simulations, and are
identical to those found in other
studies.\protect\cite{Northby,Hoare-Pal,MF} Beginning with the
pentagonal bipyramid core corresponding to  $N = 7$, the minimum energy
configuration for each succeeding cluster up to $N = 12$ can be
formally obtained by adding an atom (indicated by the shading) to a
neighboring pentagonal site to form a second five-atom ring, which is
then capped to form the $N = 13$ icosahedron. Except for $N = 17$, the
minimum energy configurations for $N = 14$ to 19 are analogous, with
succeeding clusters built up from the icosahedral core by adding an
atom to a neighboring pentagonal site to form a third five-atom ring,
which is then capped to form the $N = 19$ double icosahedron. This
correspondence between clusters $N = 8$ to 13 and $N = 14$ to 19 is
reflected in the binding energy differences, $\Delta E_b$, shown in
Fig.~\protect\ref{Fig:magic} --- the sequence for $N = 8$ to 13 is
nearly identical with the sequence for $N = 14$ to 19.
\label{Fig:clusters}}
\end{figure}

\begin{figure}
\caption{Rms bond length fluctuations as functions of temperature for $N
= 13$--15 and $N = 19$--21. The values were obtained in the same manner
as those in Fig.~\protect\ref{Fig:rms4-9}. Magic number behavior is
also evident for $N = 13$ and $N = 19$ by their significantly higher
temperatures marking the onset of the sharp rise in $\delta(T)$. The $N
= 14$ curve is somewhat unusual, showing a smaller, nearly linear rise
at about $T^* = 0.064$ until about $T^* = 0.13$ where it then undergoes
the sharp rise typical of the other clusters.
\label{Fig:rms13_19}}
\end{figure}

\begin{figure}
\caption{Minimum energy configurations for $N = 20$ to 24. The
configurations shown are the lowest energy isomers obtained from the
J-walking simulations, and are identical to those found in other
studies.\protect\cite{Northby,MF} Each configuration can be obtained
formally from the preceding one by the addition of the shaded atom.
For $N = 23$, the additional atom caps an interpenetrating icosahedral
substructure, forming a magic number structure.
\label{Fig:clusters2}}
\end{figure}

\begin{figure}
\caption{Potential energy curves as functions of temperature obtained
from J-walking simulations for clusters ranging from $N = 4$ (top
curve) to $N = 24$ (bottom curve). The magic number sizes ($N = 7$, 13,
19 and 23) are indicated by the dashed curves. Reduced units have been
used, with $V^* = V/\epsilon$ and $T^* = k_BT/\epsilon$. The dotted
lines intersecting the curves are the values for the beginning of the
transition region $T^*_\sigma$ (as characterized by a sharp rise in the
heat capacity fluctuations), and the solid-liquid transition region
heat capacity peak characteristic temperatures ($T_L^*$, $T_P^*$ and
$T_H^*$ for clusters having distinct peaks, or the inflection points
$T_L^*$ and $T_H^*$ for those clusters having heat capacity shoulders).
The transformation from the solid-like region to the liquid-like region
is evident by the change in spacing between the curves from nonuniform
at low temperatures to nearly uniform at high temperatures.
\label{Fig:potential}}
\end{figure}

\begin{figure}
\caption{Potential energy differences as functions of temperature $T^*$
and size $N$ (the vertical axis is $-\Delta \langle V(N,T)\rangle =
-[\langle V_N(T)\rangle - \langle V_{N-1}(T)\rangle]$). The temperature
mesh size is $\Delta T = 0.2$ K, or $\Delta T^* = 0.00562$. The
structure of the zero temperature curve $\Delta E_b(N)$ (which is also
shown in Fig.~\protect\ref{Fig:magic}) can be seen to extend well into
the transition regions, indicating the persistence of the solid-like
lowest energy isomer, especially for the magic number sizes. The dashed
curves correspond to the temperatures closest to the heat capacity peak
temperatures $T^*_P$ for the magic numbers $N = 13$ (long dashes), 19
(medium dashes) and 23 (short dashes). \label{Fig:DV}} \end{figure}

\begin{figure}
\caption{Normalized potential energy distribution surfaces as functions
of potential energy and temperature for $N = 13$ to 18. The surfaces
are comprised of potential energy histograms obtained from J-walking
simulations with temperature increments of $\Delta T = 0.2$ K ($\Delta
T^* = 0.00562$). In each case, the truncated peak on the left
represents the solid region and the peak on the right the dissociation
region; saddle points correspond to heat capacity peaks.
\label{Fig:surf_dis}}
\end{figure}

\begin{figure}
\caption{Normalized potential energy histograms for temperatures near
the solid-liquid transition region heat capacity peak for the magic
number clusters. The solid curves are for temperatures below $T^*_P$,
the dotted curves for temperatures above; the temperature increment is
$\Delta T = 0.1$ K ($\Delta T^* = 0.00281$). The data was obtained from
the J-walking simulations. Only the $N = 13$ distributions have a bimodal
form indicative of solid-liquid coexistence.
\label{Fig:magic_dis}}
\end{figure}

\begin{figure}
\caption{Magic number effects in cluster rms bond length fluctuations.
The plot shows the temperatures $T^*_{\delta,{\rm MC}}$, where
$\delta(T)$ obtained from Metropolis simulations begins to deviate
sharply from the nearly linear increase associated with the solid
region (see Fig.~\protect\ref{Fig:rms} for an example). For $N = 14$, 18,
22 and 24, the lower triangle signifies the temperature where
$\delta(T)$ first has a moderate rise at the end of the solid region,
and the upper triangle signifies temperature where $\delta(T)$ then
rises sharply. Also shown are the corresponding values
$T^*_{\delta,{\rm MD}}$ obtained from molecular dynamics simulations
reported in Ref.~\protect\onlinecite{BJB}; the vertical error bars
indicate the coexistence ranges found in the molecular dynamics studies.
The Metropolis values occur at much lower temperatures, but show a
similar pattern as a function of size. Also included are the heat
capacity peak temperatures $T^*_P$ and the transition ranges $\Delta T^*
= T^*_H - T^*_\sigma$, which are indicated by the dotted lines.  The
temperatures $T^*_{\delta,{\rm MC}}$ can be seen to roughly correspond
to the beginning of the transition ranges $T^*_\sigma$ for most sizes.
\label{Fig:magic_rms}}
\end{figure}

\mediumtext
\begin{table}
\caption{J-walker distributions for $4 \leq N \leq 12$. The
distributions for each cluster were generated in stages, with the
initial J-walker distribution obtained from a single Metropolis walk at
a temperature high enough to be in the cluster dissociation region, and
subsequent lower temperature distributions obtained using J-walking from
the preceding distribution; the jump attempt frequency was $P_J =
0.1$ throughout. For each distribution, $10^6$ warmup moves were made
before configurations were stored. All distributions consisted of
$10^6$ configurations stored in 10 or 50 files (depending on the amount
of computer primary memory), except for $N = 4$ and $N = 8$, which had
$1.25 \times 10^6$ and $0.75 \times 10^6$ configurations, respectively.
Configurations were stored periodically at intervals of $J_{\rm extra}$
moves. For those distributions obtained using J-walking, \%$J_A$ gives
the percentage of attempted jumps that were accepted. Absolute
temperatures are for Ne. Potential energy histograms for the $N = 11$
distributions are shown in Figure~\protect\ref{Fig:histogram}.
\label{Tbl:J-walker1}}
\begin{tabular}{*{3}{ddrd}}
\multicolumn{4}{c}{$N = 4$} &
\multicolumn{4}{c}{$N = 5$} &
\multicolumn{4}{c}{$N = 6$} \\
\cline{1-4} \cline{5-8} \cline{9-12}
$T$(K) & $T^*$ & $J_{\rm extra}$ & \%$J_A$ &
$T$(K) & $T^*$ & $J_{\rm extra}$ & \%$J_A$ &
$T$(K) & $T^*$ & $J_{\rm extra}$ & \%$J_A$ \\
\hline
15.0 & 0.421 & 400 &      &
20.0 & 0.562 & 500 &      &
15.0 & 0.421 & 600 &      \\
 8.0 & 0.225 & 400 &  8.1 &
11.0 & 0.309 & 500 & 33.4 &
12.0 & 0.337 & 600 & 43.4 \\
 4.0 & 0.112 & 400 & 12.4 &
 7.0 & 0.197 & 500 &  4.4 &
 8.0 & 0.225 & 600 &  5.0 \\
     &       &      &      &
     &       &      &      &
 4.0 & 0.112 & 600 &  8.6 \\
\\
\multicolumn{4}{c}{$N = 7$} &
\multicolumn{4}{c}{$N = 8$} &
\multicolumn{4}{c}{$N = 9$} \\
\cline{1-4} \cline{5-8} \cline{9-12}
$T$(K) & $T^*$ & $J_{\rm extra}$ & \%$J_A$ &
$T$(K) & $T^*$ & $J_{\rm extra}$ & \%$J_A$ &
$T$(K) & $T^*$ & $J_{\rm extra}$ & \%$J_A$ \\
\hline
15.0 & 0.421 & 1000 &      &
20.0 & 0.562 &  800 &      &
20.0 & 0.562 &  900 &      \\
12.0 & 0.337 & 1000 & 18.2 &
15.0 & 0.421 &  800 & 45.9 &
15.0 & 0.421 &  900 & 28.1 \\
 8.0 & 0.225 & 1000 &  9.1 &
12.0 & 0.337 &  800 &  9.3 &
11.5 & 0.323 &  900 &  7.1 \\
     &       &      &      &
 8.0 & 0.225 &  800 & 11.1 &
 7.0 & 0.197 &  900 &  6.2 \\
\\
\multicolumn{4}{c}{$N = 10$} &
\multicolumn{4}{c}{$N = 11$} &
\multicolumn{4}{c}{$N = 12$} \\
\cline{1-4} \cline{5-8} \cline{9-12}
$T$(K) & $T^*$ & $J_{\rm extra}$ & \%$J_A$ &
$T$(K) & $T^*$ & $J_{\rm extra}$ & \%$J_A$ &
$T$(K) & $T^*$ & $J_{\rm extra}$ & \%$J_A$ \\
\hline
20.0 & 0.562 & 1000 &      &
20.0 & 0.562 & 1000 &      &
20.0 & 0.562 & 1200 &      \\
16.0 & 0.449 & 1000 & 34.2 &
16.0 & 0.449 & 1100 & 19.8 &
17.0 & 0.478 & 1200 & 30.8 \\
12.0 & 0.337 & 1000 &  4.4 &
12.0 & 0.337 & 1100 &  6.2 &
14.0 & 0.393 & 1200 & 13.1 \\
 8.0 & 0.225 & 1000 &  9.6 &
 8.0 & 0.225 & 1100 &  7.6 &
10.0 & 0.281 & 1200 &  7.2 \\
 5.0 & 0.140 & 1000 &  6.5 &
 5.0 & 0.140 & 1100 &  5.9 &
 7.0 & 0.197 & 1200 &  8.6 \\
     &       &      &      &
     &       &      &      &
 4.0 & 0.112 & 1200 &  6.5 \\
\end{tabular}
\end{table}

\newpage
\widetext
\begin{table}
\caption{J-walker distributions for $13 \leq N \leq 24$. The distributions
were obtained as described in Table~\protect\ref{Tbl:J-walker1}. Potential
energy histograms for the $N = 19$ distributions are shown in
Figure~\protect\ref{Fig:histogram}.
\label{Tbl:J-walker2}}
\squeezetable
\begin{tabular}{*{4}{ddrd}}
\multicolumn{4}{c}{$N = 13$} &
\multicolumn{4}{c}{$N = 14$} &
\multicolumn{4}{c}{$N = 15$} &
\multicolumn{4}{c}{$N = 16$} \\
\cline{1-4} \cline{5-8} \cline{9-12} \cline{13-16}
$T$(K) & $T^*$ & $J_{\rm extra}$ & \%$J_A$ &
$T$(K) & $T^*$ & $J_{\rm extra}$ & \%$J_A$ &
$T$(K) & $T^*$ & $J_{\rm extra}$ & \%$J_A$ &
$T$(K) & $T^*$ & $J_{\rm extra}$ & \%$J_A$ \\
\hline
22.0 & 0.618 & 1300 &      &
22.0 & 0.618 & 1400 &      &
25.0 & 0.702 & 1500 &      &
25.0 & 0.702 &  800 &      \\
18.0 & 0.506 & 1300 & 28.2 &
18.0 & 0.506 & 1400 & 18.3 &
21.0 & 0.590 & 1500 & 49.6 &
20.0 & 0.562 & 1600 & 23.0 \\
14.0 & 0.393 & 1300 &  5.0 &
14.0 & 0.393 & 1400 &  6.0 &
17.5 & 0.492 & 1500 & 11.6 &
16.5 & 0.463 & 1600 &  9.0 \\
10.5 & 0.295 & 1300 &  8.4 &
10.0 & 0.281 & 1400 &  3.8 &
14.0 & 0.393 & 1500 & 11.2 &
13.0 & 0.463 & 1600 & 11.5 \\
 7.0 & 0.197 & 1300 &  4.3 &
 6.0 & 0.169 & 1400 &  2.8 &
11.0 & 0.309 & 1500 & 13.5 &
10.0 & 0.281 &  800 & 12.3 \\
 4.0 & 0.112 &  130 &  6.5 &
 3.0 & 0.084 &  140 &  2.0 &
 8.0 & 0.225 &  150 &  8.5 &
 7.0 & 0.197 &  400 &  6.0 \\
 2.0 & 0.056 &  130 &  3.6 &
     &       &      &      &
 6.0 & 0.169 &  150 & 20.5 &
 4.5 & 0.126 &  160 &  7.3 \\
     &       &      &      &
     &       &      &      &
 4.0 & 0.112 &  150 & 13.5 &
 2.5 & 0.070 &  160 &  3.9 \\
     &       &      &      &
     &       &      &      &
 2.0 & 0.056 &  150 &  2.2 &
     &       &      &      \\
\\
\multicolumn{4}{c}{$N = 17$} &
\multicolumn{4}{c}{$N = 18$} &
\multicolumn{4}{c}{$N = 19$} &
\multicolumn{4}{c}{$N = 20$} \\
\cline{1-4} \cline{5-8} \cline{9-12} \cline{13-16}
$T$(K) & $T^*$ & $J_{\rm extra}$ & \%$J_A$ &
$T$(K) & $T^*$ & $J_{\rm extra}$ & \%$J_A$ &
$T$(K) & $T^*$ & $J_{\rm extra}$ & \%$J_A$ &
$T$(K) & $T^*$ & $J_{\rm extra}$ & \%$J_A$ \\
\hline
25.0 & 0.702 &  850 &      &
25.0 & 0.702 &  900 &      &
25.0 & 0.702 &  950 &      &
25.0 & 0.702 & 1000 &      \\
21.0 & 0.590 & 1700 & 33.9 &
21.0 & 0.590 & 1800 & 26.0 &
21.0 & 0.590 & 1900 & 20.8 &
21.0 & 0.590 & 2000 & 16.6 \\
18.0 & 0.506 & 1700 & 15.3 &
18.0 & 0.506 & 1800 & 15.8 &
17.5 & 0.492 & 1900 & 10.7 &
17.5 & 0.492 & 2000 & 11.3 \\
15.0 & 0.421 & 1700 & 17.7 &
15.0 & 0.421 & 1800 & 17.8 &
14.0 & 0.393 & 1900 & 11.6 &
14.0 & 0.393 & 2000 & 11.3 \\
12.0 & 0.337 &  850 & 16.4 &
12.0 & 0.337 &  900 & 15.8 &
11.0 & 0.309 & 1900 & 12.2 &
11.0 & 0.309 & 2000 & 11.2 \\
 9.0 & 0.253 &  850 &  9.3 &
 9.0 & 0.253 &  900 &  7.6 &
 9.0 & 0.253 &  950 & 14.0 &
 9.0 & 0.253 & 1000 & 13.9 \\
 6.5 & 0.183 &  170 &  9.5 &
 6.5 & 0.183 &  900 &  9.2 &
 6.5 & 0.183 &  950 &  9.3 &
 6.5 & 0.183 & 1000 &  9.6 \\
 4.5 & 0.126 &  170 & 13.4 &
 4.0 & 0.112 &  180 &  4.6 &
 4.0 & 0.112 &  190 &  5.1 &
 4.5 & 0.126 &  200 & 11.9 \\
 3.0 & 0.084 &  170 & 13.4 &
 2.5 & 0.070 &  180 &  7.1 &
 2.5 & 0.070 &  190 &  7.4 &
 3.0 & 0.084 &  200 & 10.7 \\
 1.5 & 0.042 &  170 &  1.5 &
 1.5 & 0.042 &  180 &  6.0 &
 1.5 & 0.042 &  190 &  6.0 &
 2.0 & 0.056 &  200 & 11.9 \\
     &       &      &      &
     &       &      &      &
     &       &      &      &
 1.0 & 0.028 &  200 &  1.0 \\
\\
\multicolumn{4}{c}{$N = 21$} &
\multicolumn{4}{c}{$N = 22$} &
\multicolumn{4}{c}{$N = 23$} &
\multicolumn{4}{c}{$N = 24$} \\
\cline{1-4} \cline{5-8} \cline{9-12} \cline{13-16}
$T$(K) & $T^*$ & $J_{\rm extra}$ & \%$J_A$ &
$T$(K) & $T^*$ & $J_{\rm extra}$ & \%$J_A$ &
$T$(K) & $T^*$ & $J_{\rm extra}$ & \%$J_A$ &
$T$(K) & $T^*$ & $J_{\rm extra}$ & \%$J_A$ \\
\hline
25.0 & 0.702 & 1050 &      &
25.0 & 0.702 & 1100 &      &
25.0 & 0.702 & 1150 &      &
25.0 & 0.702 & 1200 &      \\
21.0 & 0.590 & 2100 & 13.6 &
21.0 & 0.590 & 2200 & 11.9 &
21.0 & 0.590 & 2300 & 10.8 &
22.0 & 0.618 & 2400 & 23.7 \\
17.5 & 0.492 & 2100 & 11.7 &
17.5 & 0.492 & 2200 & 11.8 &
18.0 & 0.506 & 2300 & 18.3 &
19.0 & 0.534 & 2400 & 18.7 \\
14.0 & 0.393 & 2100 & 10.9 &
14.0 & 0.393 & 2200 & 10.5 &
15.0 & 0.421 & 2300 & 17.1 &
16.0 & 0.449 & 2400 & 16.9 \\
11.0 & 0.309 & 2100 & 10.0 &
11.0 & 0.309 & 1100 &  9.6 &
12.5 & 0.351 & 2300 & 20.2 &
13.0 & 0.365 & 2400 & 14.0 \\
 8.5 & 0.239 & 1050 &  8.4 &
 8.5 & 0.239 & 1100 &  7.4 &
10.5 & 0.295 & 2300 & 17.2 &
10.5 & 0.295 & 2400 & 11.3 \\
 6.0 & 0.169 & 1050 &  7.9 &
 6.0 & 0.169 & 1100 &  7.4 &
 8.5 & 0.239 & 1150 & 13.0 &
 8.5 & 0.239 & 1200 & 12.5 \\
 4.0 & 0.112 &  210 &  8.1 &
 4.0 & 0.112 &  220 &  7.3 &
 6.5 & 0.183 &  230 & 16.4 &
 6.5 & 0.183 & 1200 & 15.0 \\
 2.5 & 0.070 &  210 &  5.9 &
 2.5 & 0.070 &  220 &  5.1 &
 4.5 & 0.126 &  230 &  9.9 &
 4.5 & 0.126 &  240 &  8.7 \\
 1.5 & 0.042 &  210 &  4.6 &
 1.5 & 0.042 &  220 &  4.1 &
 3.0 & 0.084 &  230 &  8.5 &
 3.0 & 0.084 &  240 &  7.7 \\
     &       &      &      &
     &       &      &      &
 2.0 & 0.056 &  230 &  9.3 &
 2.0 & 0.056 &  240 &  8.6 \\
     &       &      &      &
     &       &      &      &
 1.0 & 0.028 &  230 &  0.5 &
 1.0 & 0.028 &  240 &  0.4 \\
\end{tabular}
\end{table}

\newpage
\widetext
\begin{table}
\caption{Solid-liquid transition region heat capacity peak parameters for
$4 \leq N \leq 24$. These values were obtained by smoothing and
interpolating the J-walking data shown in Figs.~\protect\ref{Fig:Cv1}
and \protect\ref{Fig:Cv2}.  The points $\left(T^*_P,\langle
C_V^*\rangle_P/N\right)$ are the heat capacity peaks, while the points
$\left(T^*_H,\langle C_V^*\rangle_H/N\right)$ are the local minima
occurring on the high temperature side of each  peak. For those clusters
having heat capacity shoulders instead of peaks ($N = 7$, 15--17), the
points $\left(T^*_H,\langle C_V^*\rangle_H/N\right)$ are the inflection
points.  The points $\left(T^*_L,\langle C_V^*\rangle_L/N\right)$ are the
inflection points on the low temperature side of the peak. The uncertainty
estimates are standard deviations of the values obtained using different
curve smoothing parameters.
\label{Tbl:J-walker3}}
\squeezetable
\begin{tabular}{r*{3}{dddd}}
N &
\multicolumn{2}{c}{$T^*_L$} &
\multicolumn{2}{c}{$\langle C_V^*\rangle_L/N$} &
\multicolumn{2}{c}{$T^*_P$} &
\multicolumn{2}{c}{$\langle C_V^*\rangle_P/N$} &
\multicolumn{2}{c}{$T^*_H$} &
\multicolumn{2}{c}{$\langle C_V^*\rangle_H/N$} \\
\hline
4 & & & & & & & & & & & & \\
5 & & & & & & & & & & & & \\
6 & 0.0652 & $\pm$0.0004 &
3.112 & $\pm$0.008 &
0.1036 & $\pm$0.0008 &
3.595 & $\pm$0.001 &
0.1344 & $\pm$0.0003 &
3.540 & $\pm$0.002 \\
7 & 0.125 & $\pm$0.002 &
3.61 & $\pm$0.04 &
 &  &
 &  &
0.190 & $\pm$0.004 &
4.39 & $\pm$0.01 \\
8 & & & & & & & & & & & & \\
9 & 0.137 & $\pm$0.002 &
3.81 & $\pm$0.04 &
0.1943 & $\pm$0.0006 &
4.470 & $\pm$0.002 &
0.208 & $\pm$0.001 &
4.455 & $\pm$0.002 \\
10 & 0.145 & $\pm$0.002 &
3.99 & $\pm$0.06 &
0.1942 & $\pm$0.0008 &
4.834 & $\pm$0.001 &
0.240 & $\pm$0.002 &
4.687 & $\pm$0.001 \\
11 & 0.162 & $\pm$0.002 &
4.18 & $\pm$0.05 &
0.215 & $\pm$0.001 &
5.010 & $\pm$0.002 &
0.255 & $\pm$0.001 &
4.925 & $\pm$0.003 \\
12 & 0.224 & $\pm$0.004 &
5.0 & $\pm$0.2 &
0.272 & $\pm$0.002 &
6.326 & $\pm$0.003 &
0.317 & $\pm$0.001 &
6.146 & $\pm$0.001 \\
13 & 0.2593 & $\pm$0.0006 &
7.10 & $\pm$0.06 &
0.2875 & $\pm$0.0007 &
9.06 & $\pm$0.01 &
0.3553 & $\pm$0.0008 &
6.794 & $\pm$0.004 \\
14 & 0.263 & $\pm$0.002 &
5.8 & $\pm$0.1 &
0.307 & $\pm$0.001 &
7.124 & $\pm$0.002 &
0.356 & $\pm$0.001 &
6.770 & $\pm$0.002 \\
15 & 0.265 & $\pm$0.002 &
5.52 & $\pm$0.04 &
 &  &
 &  &
0.339 & $\pm$0.002 &
6.589 & $\pm$0.007 \\
16 & 0.194 & $\pm$0.001 &
4.29 & $\pm$0.03 &
 &  &
 &  &
0.321 & $\pm$0.003 &
5.86 & $\pm$0.02 \\
17 & 0.209 & $\pm$0.004 &
4.5 & $\pm$0.1 &
 &  &
 &  &
0.295 & $\pm$0.002 &
5.538 & $\pm$0.006 \\
18 & 0.224 & $\pm$0.002 &
4.80 & $\pm$0.06 &
0.274 & $\pm$0.001 &
5.834 & $\pm$0.001 &
0.326 & $\pm$0.002 &
5.652 & $\pm$0.002 \\
19 & 0.243 & $\pm$0.001 &
5.92 & $\pm$0.08 &
0.2739 & $\pm$0.0006 &
7.318 & $\pm$0.002 &
0.352 & $\pm$0.002 &
5.794 & $\pm$0.005 \\
20 & 0.2519 & $\pm$0.0004 &
5.77 & $\pm$0.02 &
0.2858 & $\pm$0.0003 &
7.085 & $\pm$0.001 &
0.3610 & $\pm$0.0008 &
5.756 & $\pm$0.001 \\
21 & 0.257 & $\pm$0.002 &
5.37 & $\pm$0.06 &
0.300 & $\pm$0.002 &
6.383 & $\pm$0.005 &
0.371 & $\pm$0.001 &
5.843 & $\pm$0.001 \\
22 & 0.258 & $\pm$0.003 &
5.5 & $\pm$0.1 &
0.298 & $\pm$0.001 &
6.333 & $\pm$0.002 &
0.3727 & $\pm$0.0009 &
5.726 & $\pm$0.001 \\
23 & 0.265 & $\pm$0.001 &
6.05 & $\pm$0.07 &
0.297 & $\pm$0.001 &
7.383 & $\pm$0.003 &
0.386 & $\pm$0.001 &
5.764 & $\pm$0.001 \\
24 & 0.257 & $\pm$0.001 &
5.56 & $\pm$0.06 &
0.2917 & $\pm$0.0002 &
6.703 & $\pm$0.001 &
0.3787 & $\pm$0.0006 &
5.557 & $\pm$0.007 \\
\end{tabular}
\end{table}


\begin{references}
\bibitem{acceptance} Distributions were typically used until the jump
    acceptance dropped down to about 10\%, although the values were often
    greater for the wider distributions in the high temperature dissociation
    region and lower for the narrower distributions in the low
    temperature solid region.
\bibitem{computers} Calculations for cluster sizes $N = 4$ to 8 were
    performed on a DECstation 5000/200 UNIX workstation equipped with 16 Mb
    of RAM. The memory was upgraded to 64 Mb for clusters $N = 9$ to 16, and
    then upgraded again to 96 Mb for clusters $N = 17$ to 19. Calculations
    for cluster sizes $N = 20$, 22 and 24 were performed on a DEC Alpha
    3000/500 AXP equipped with 196 Mb of RAM. Calculations for cluster sizes
    $N = 21$ and 23 were performed on a DEC Alpha 3000/300L AXP equipped with
    32 Mb of RAM.
\bibitem{Sav_Gol} The Savitsky-Golay algorithm was used for the data
    smoothing, and 2D cubic splines were used for the interpolation. These
    routines are implemented in the MS-DOS graphics software package Axum,
    version 3.0. Different fits were obtained by using different smoothing
    orders, corresponding to 7, 9, 11 or 13 surrounding data points being
    used to fit each point.
\bibitem{diffusion} The self diffusion constants as functions of temperature
    differed substantially from those obtained from molecular dynamics
    calculations by Beck and Marchioro (Ref.~\onlinecite{BM}) by their lack
    of a sharp rise in the transition region. A subsequent instantaneous
    normal mode analysis by Adams and Stratt (Ref.~\onlinecite{AS2}) using
    an expression for the diffusion constant that was more sensitive to the
    low frequency part of the phonon spectrum gave results that were in
    agreement with the molecular dynamics calculations. The curves for
    Ar$_7$ and Ar$_8$ remained very much alike.
\bibitem{min_config} Because the J-walking runs were started at high
    temperatures from random configurations, and the temperatures were
    decreased in small increments down to zero, the simulations also served
    as global minimization calculations, akin to simulated annealing. In all
    cases, the lowest energy configuration obtained with J-walking was the
    same as that reported in Refs.~\onlinecite{Northby} and \onlinecite{MF}.
\end{references}
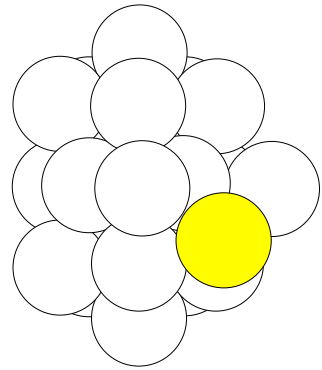
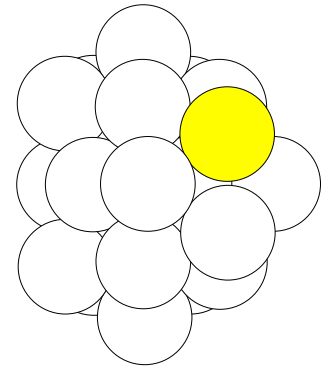
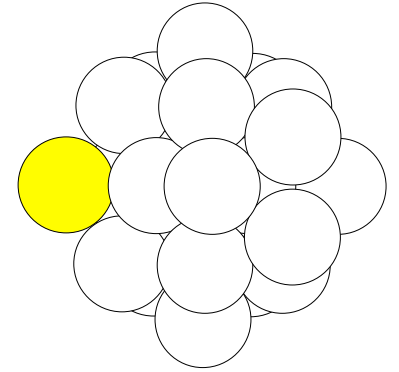
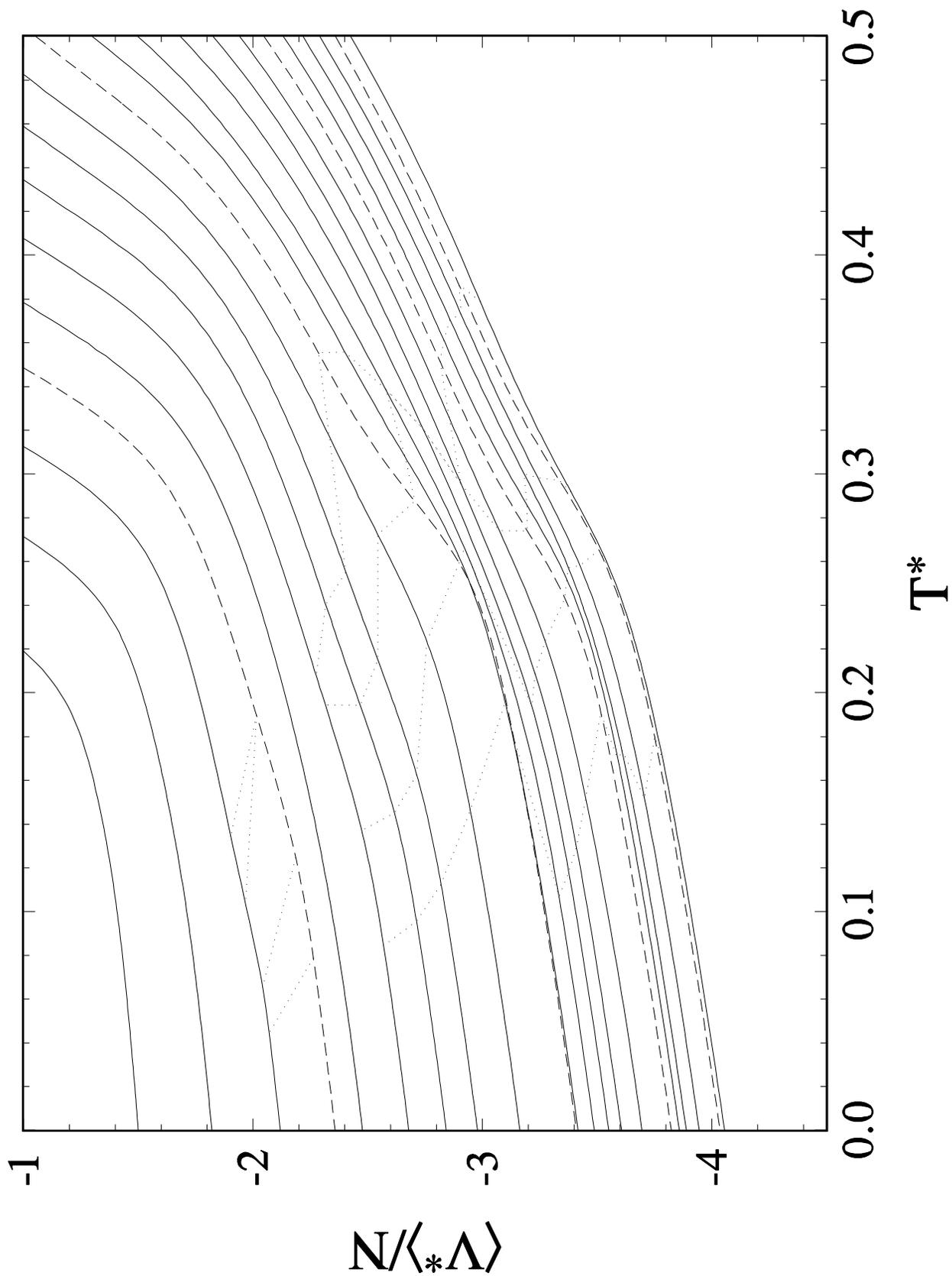
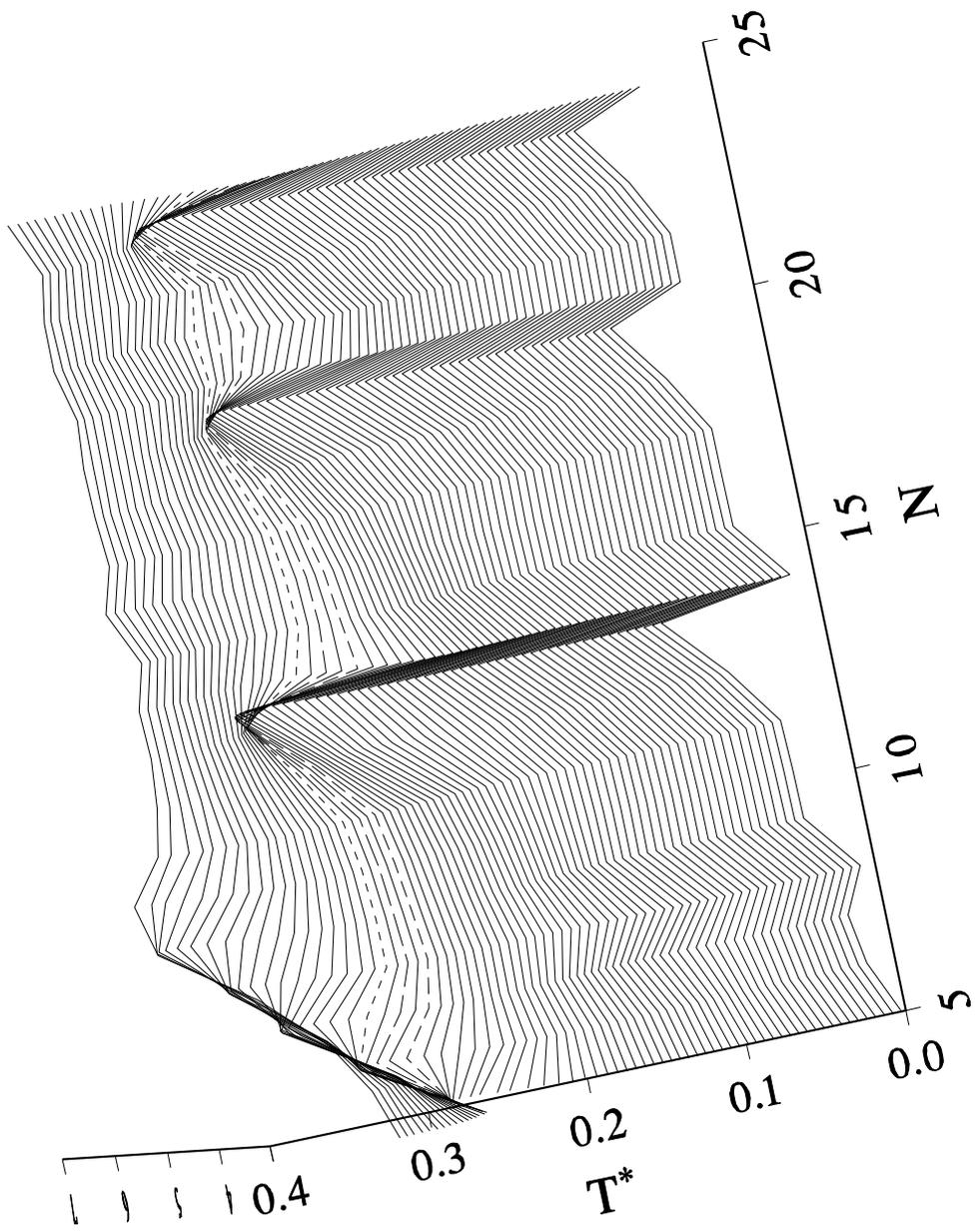
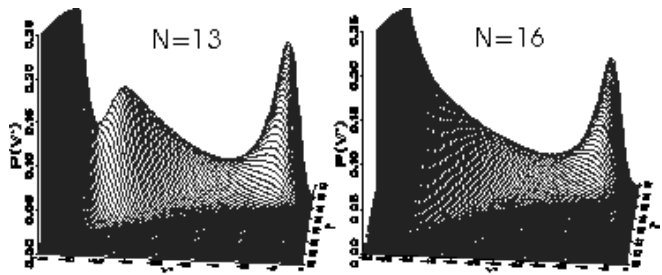
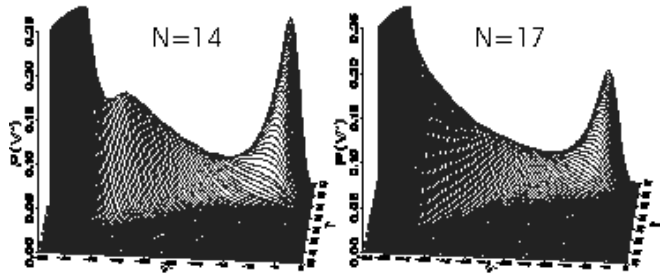
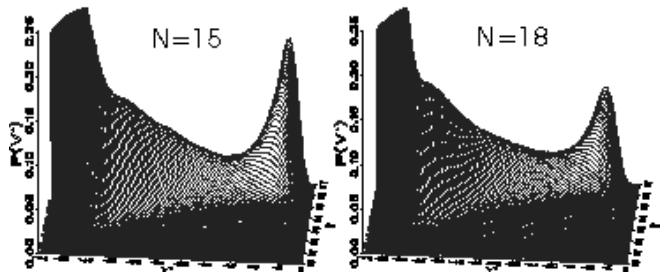
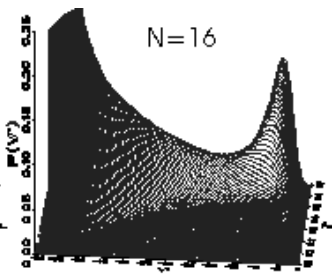
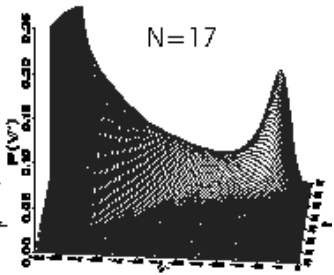
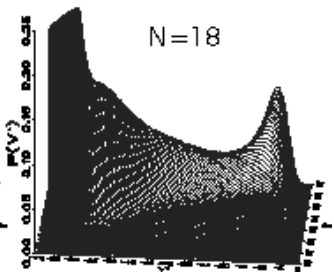
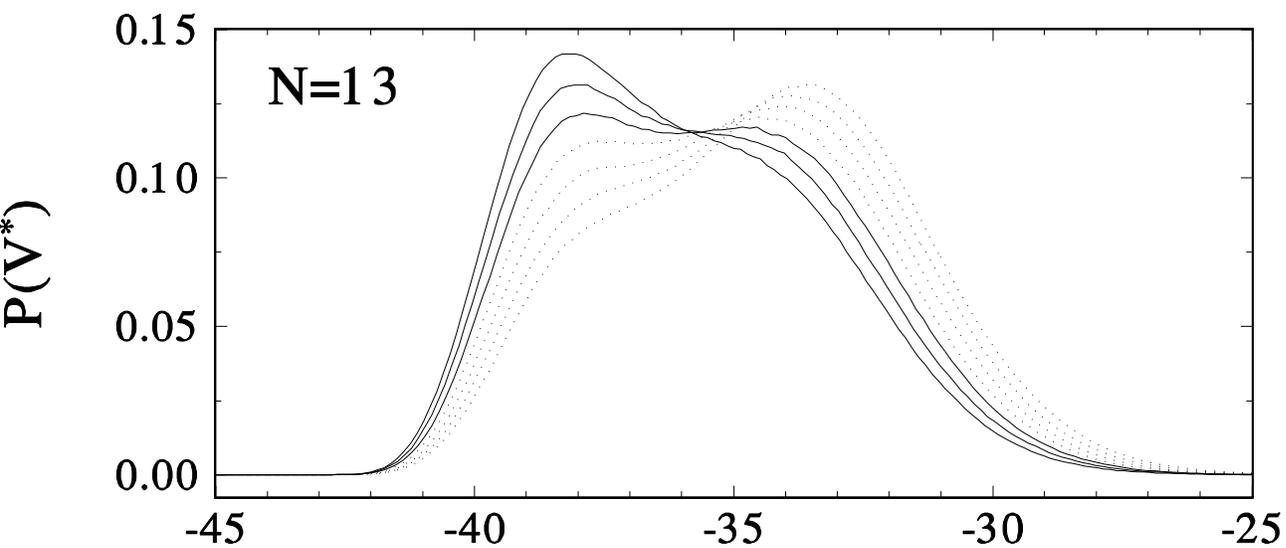
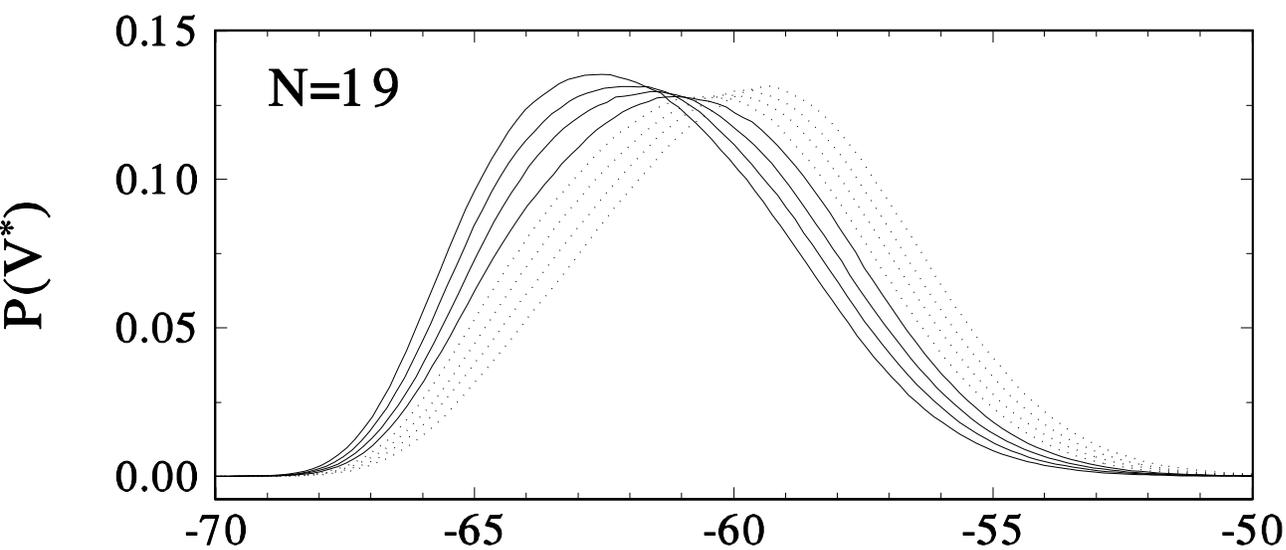
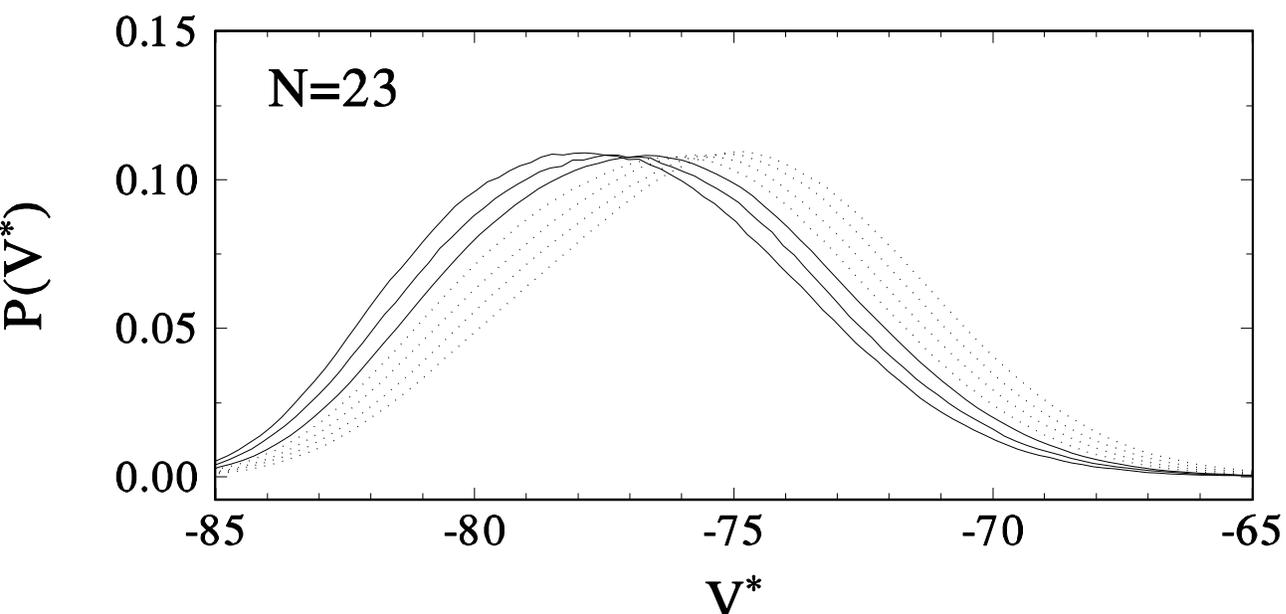
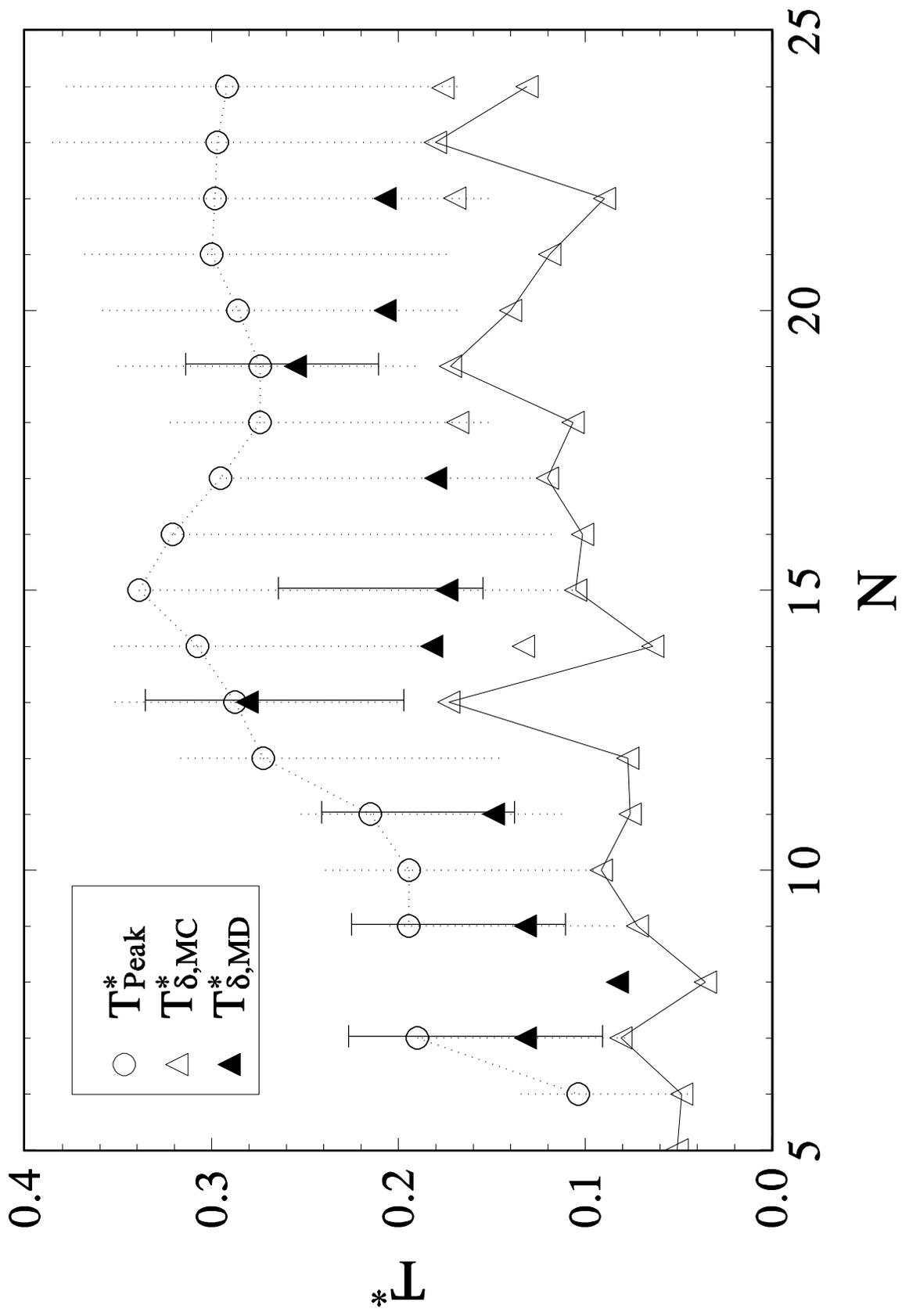
\end{document}